\documentclass{jfm}
\usepackage{graphicx}
\usepackage{caption}
\usepackage{subcaption}
\graphicspath{ {./images/} }
\usepackage{xcolor}
\usepackage{epstopdf, epsfig}
\usepackage{multirow, booktabs}
\usepackage{amsfonts}
\usepackage{amsmath}
\usepackage{setspace} 
\usepackage{bm} 
\usepackage{enumitem}
\usepackage{xstring}
\usepackage{url}

\usepackage{cleveref} 

\makeatletter
\def\tagform@#1{%
  \maketag@@@{(#1\unskip\@@italiccorr)}%
}
\makeatother

\crefname{equation}{equation}{equations}
\Crefname{equation}{Equation}{Equations}

\shorttitle{Liquid bridges between horizontal cylinders}
\shortauthor{A.~J.~Bok\'anyi-T\'oth et al.}

\title{Liquid bridges between horizontal cylinders: effects of gravity and electric fields}

\author{Agnes J.~Bok\'anyi-T\'oth\aff{1}
  \corresp{\email{A.J.Bokanyi-Toth@lboro.ac.uk}},
   Andrew~J.~Archer\aff{1}\corresp{\email{A.J.Archer@lboro.ac.uk}},
   Radu Cimpeanu\aff{2}\corresp{\email{Radu.Cimpeanu@warwick.ac.uk}},
 H.~C.~Hemaka~Bandulasena\aff{3}\corresp{\email{H.C.H.Bandulasena@lboro.ac.uk}},
 Gyula I.~T\'oth\aff{1}\corresp{\email{G.I.Toth@lboro.ac.uk}}\\
 \and Dmitri Tseluiko\aff{1}\corresp{\email{D.Tseluiko@lboro.ac.uk}}}

\affiliation{\aff{1}Department of Mathematical Sciences and Interdisciplinary Centre for Mathematical Modelling, Loughborough University,
Loughborough, LE11 3TU, UK
\aff{2}Mathematics Institute, University of Warwick, Coventry, CV4 7AL, UK
\aff{3}Department of Chemical Engineering, Loughborough University,
Loughborough, LE11 3TU, UK}

\begin{document}

\maketitle

\begin{abstract}
\looseness=-1
Liquid bridges suspended between two parallel horizontal cylinders are studied using experiments, reduced-order mathematical modelling and numerical simulations. Both non-electrified and electrified configurations are considered, with the cylinders acting as electrodes between which a potential difference can be applied. The initial focus is on equilibrium bridge shapes, while the dynamics is examined only in the non-electrified case. In the experiments, transformer-oil bridges are investigated and modelled as perfect dielectrics. The electric field counteracts the gravitational effects and causes the bridges to rise and become flatter, with shapes that are well described by Young--Laplace-type equations containing non-local electric-field contributions evaluated using a boundary-element method. Bifurcation diagrams of these solutions are constructed by pseudo-arclength continuation to characterise the dependence of bridge shapes on liquid volume and electric-field strength. In the absence of an electric field, the transient relaxation to equilibrium is analysed using a reduced-order model developed using Onsager's variational principle, with comparison to direct numerical simulations. The steady states predicted by the reduced-order model agree closely with those of the full formulation over a wide parameter range, and the dynamics is well captured in the overdamped regime. 
\end{abstract}


\section{Introduction}

The behaviour of liquids in contact with solid surfaces is strongly influenced by geometry, and can become significantly more complex in the presence of curvature, edges or confinement. One important example 
is provided by liquid bridges, which arise when a volume of liquid is deposited into the gap between nearby solid bodies. Understanding such systems is important for a wide range of applications, including granular materials, lubrication, porous media and foams. Such bridges have been studied in a wide range of geometrical configurations, including between planar, spherical and cylindrical surfaces. A number of classical works have provided detailed descriptions of equilibrium capillary-bridge shapes using analytical and semi-analytical approaches based on the Young--Laplace equation, particularly in idealised geometries, see e.g.\ \cite{rose1958, cross1963, melrose1966, clark1968, orr1975}. A general theoretical framework for capillary phenomena, including liquid bridges in various geometries, is provided in the monograph by \cite{myshkis1987}, with particular emphasis on situations in which gravitational effects are weak or negligible, in contrast to the regime considered in the present study.  A substantial body of work has been devoted to understanding the capillary forces exerted by liquid menisci, including liquid bridges -- see, for example, the comprehensive review by \citet{Capillary_forces}.

Later, the effect of gravity on liquid bridges was analysed in more detail, including its influence on equilibrium shapes and their stability, see, for example, \citet{lowry1995, laveron1995, adams2002}. Gravitational effects can typically be neglected when the size of the bridge is small compared to the capillary length, but become important when these length scales are comparable. Although multiple equilibrium solutions may exist, only stable configurations are physically realised, and changes in stability govern transitions between different bridge shapes. The stability of capillary surfaces, including liquid bridges, is reviewed by \citet{bostwick2015}, and is closely related to the bifurcation structure of solutions examined in our study.

A configuration of primary interest in the present work is that of liquid bridges between two parallel identical horizontal cylinders. This problem was first studied in detail by \cite{PRINCEN}, who analysed this problem in the absence of gravity and in the limit of sufficiently long bridges, for which the bridge cross-section becomes nearly uniform away from the end menisci. In this regime, adding liquid increases the length of the bridge without significantly affecting its cross-sectional shape, so that a two-dimensional (2D) approximation applies. Three-dimensional (3D) extensions to non-parallel cylinders, for example, have also been considered \cite[see e.g.][]{virozub2009}. When gravitational effects are taken into account, a closely related 2D configuration was analysed by \citet{Cooray2016}, who considered steady-state liquid bridges between parallel horizontal cylinders in the presence of gravity. They derived analytical descriptions of the liquid--air interface based on exact solutions of the Young--Laplace equation involving incomplete elliptic integrals, and obtained approximations in the limits of small and large liquid volumes. They also investigated the maximum trapping capacity between the cylinders, showing that the maximum cross-sectional area is achieved when the separation distance is approximately twice the capillary length.

This motivates 
the present study, where we investigate how the application of an electric field, in addition to gravity, modifies the behaviour of liquid bridges. In particular, we consider liquid bridges formed between pairs of identical horizontal cylinders, using transformer oils that can be modelled as perfect dielectrics, with the cylinders acting as electrodes between which a potential difference can be applied. Electric fields can significantly affect the behaviour of liquid interfaces through Maxwell stresses. A wide range of such phenomena has been studied, including interfacial deformation, instabilities and electrically driven flows, see, for example, the reviews by \citet{vlahovska2019electrohydrodynamics, Papageorgiou_2019} and the monograph by \citet{Castellanos}. {Within this broad area, previous studies of electrified liquid bridges have focused primarily on the axisymmetric configuration of liquid bridges between parallel planar electrodes with pinned three-phase contact lines, analysing mainly equilibrium shapes, their stability and the associated bifurcation structure under axial electric fields, both in the absence and presence of axial gravity, 
see e.g.\ \citet{gonzalez_etal_1989,ramos_castellanos_1993,volkov_etal_2005} and the review by \cite{Montanero_PonceTorres_2020}. Extending the analysis to the current setup of liquid bridges between horizontal cylinders requires addressing several additional challenges because such bridges are inherently non-axisymmetric, the contact lines are free to move over the surfaces of the cylinders, and the bridge behaviour is governed by the coupled effects of the curvature of the supporting cylinders, gravity, and the non-local electric field.}

We begin by considering equilibrium configurations, which constitute the initial focus of the present work. For such states, the flow is absent, and the corresponding liquid--air interfaces are determined from Young--Laplace-type equations incorporating non-local electric-field contributions, in a manner similar to that employed for axisymmetric electrified liquid bridges \cite[e.g.][]{ramos_castellanos_1993,volkov_etal_2005}. To evaluate non-local electric-field contributions, we use a boundary-element method. 
{The time-dependent behaviour of liquid bridges is of independent interest, as it can involve non-trivial evolution and regime transitions, and it has been widely studied, particularly in axisymmetric geometries \citep[see, e.g.,][]{suryo_basaran_2006,slobozhanin_etal_2012,li2016}. 
However, a complete time-dependent electrohydrodynamic formulation for the present configuration requires solving simultaneously for the fluid motion and the electric field, which are coupled through the evolving free surfaces. This leads to a considerably more involved moving-boundary problem, particularly in the presence of moving contact lines on curved solid surfaces, and is beyond the scope of the present study.} {Nevertheless, 
the transient dynamics of non-electrified liquid bridges is of considerable interest in its own right and provides a natural first step towards understanding the evolution of liquid bridges between horizontal cylinders.} {Accordingly, we derive a reduced-order dynamical model for non-electrified liquid bridges between horizontal cylinders using Onsager's variational principle \cite[see e.g.][]{DoiSoft,DoiMan,LopesThiele}. To our knowledge, this is the first reduced-order dynamical model for liquid bridges between horizontal cylinders derived within Onsager's variational framework.
We validate its predictions against} direct numerical simulations (DNS) performed within the open-source volume-of-fluid (VOF) platform Basilisk \citep{popinet2009accurate, popinet2015quadtree}.

The rest of the paper is organised as follows. In \S~\ref{sec:2}, we describe the experimental setup, and in \S~\ref{sec:3} the governing equations. \S~\ref{sec:4} presents steady-state solutions, including their mathematical formulation, computation and comparison with experiments. \S~\ref{sec:dynamics} studies the dynamics of liquid bridges in the absence of an electric field, including the derivation of a reduced-order Onsager model, analysis of its predicted dynamics and comparison with DNS. Details of our DNS implementation within the Basilisk platform are provided in Appendix~\ref{sec:DNS}. Finally, \S~\ref{sec:10} presents the conclusions.

\section{Experiments}
\label{sec:2}

\begin{figure}
    \centering
    \includegraphics[width=1.0\textwidth]{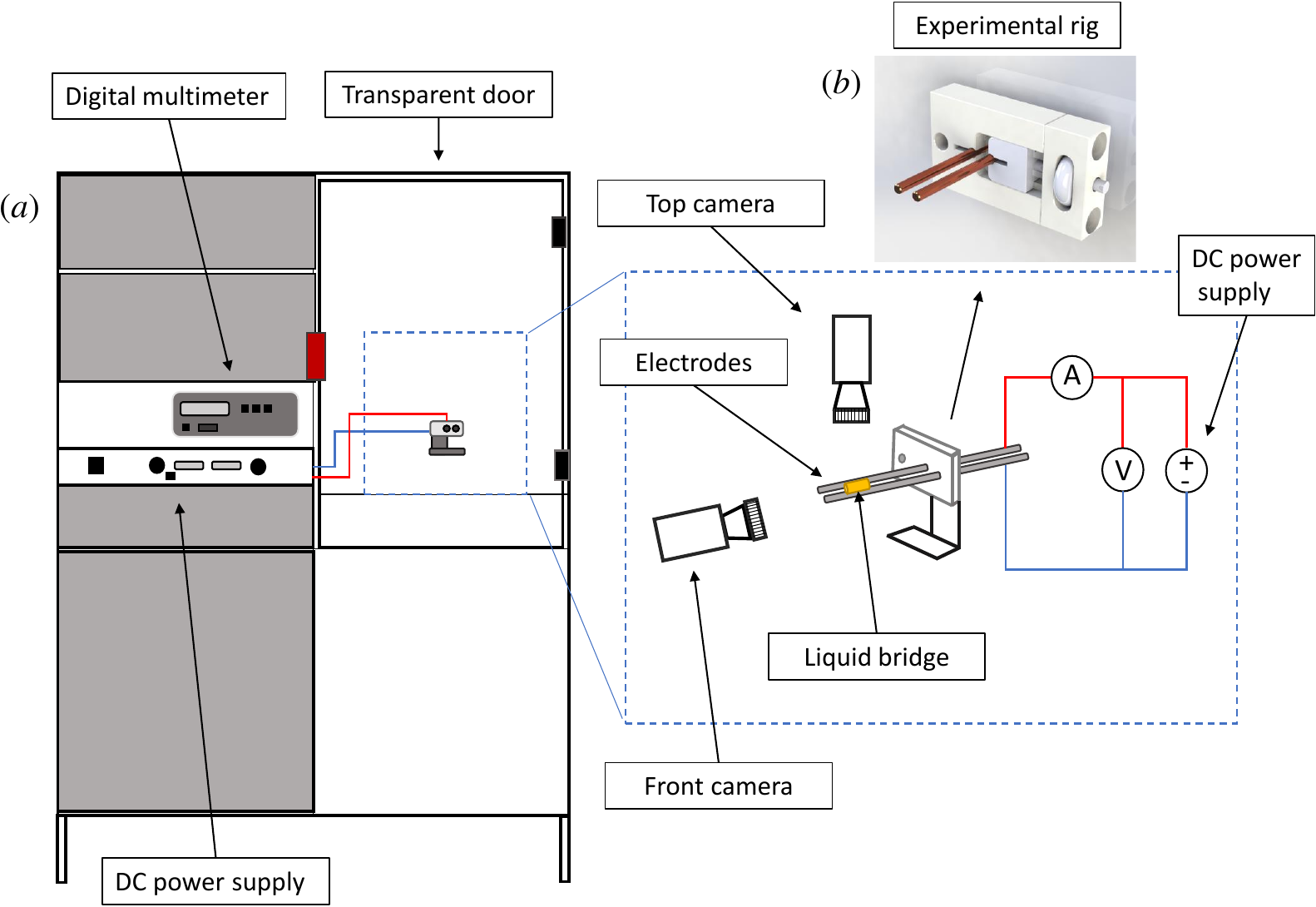}
    \caption{Schematic diagram of (\textit{a}) the safety box, including (\textit{b}) the experimental rig, DC power supply, digital multimeter, top and front view cameras.}
    \label{schematic_picture}
\end{figure}

To study liquid bridges suspended between two horizontal cylinders under an applied electric field, we designed a customised experimental setup (see figure~\ref{schematic_picture}). Two parallel cylindrical electrodes with a $2$ mm diameter are held by an adjustable plastic rig, where the separation distance can be varied (figure~\ref{schematic_picture}\textit{b}). The electrodes are connected to a high-voltage DC power source (model PS/FJ30R04.0-22, FJ Glassman High Voltage Inc.), allowing the application of potentials of up to $4\,000$ V, and the system is enclosed in a transparent box with a safety interlock (figure~\ref{schematic_picture}\textit{a}). A digital multimeter (Philips PM 2522A VA${\mathrm\Omega}$) is used to measure potential difference. 
The electrodes can be replaced, in order to vary properties such as the contact angle with the liquid, and were made of platinised titanium, copper and stainless steel rods.

\begin{table}
        \caption{Parameters of the transformer oils used in the experiments.} 
        	\label{tab:Para_Oils}
	\centering
	\begin{tabular}{l l l l l l l}
		\toprule
        \textbf{Parameter}  & \textbf{Unit}  & \textbf{ Mineral oil} & \textbf{Castor oil} & \textbf{Silicone oil} & \textbf{Silicone oil} \\ 
          &  &  & & \textbf{ $5$ $\mathrm{cSt}$} & \textbf{$500$  $\mathrm{cSt}$} \\ 
		\midrule
	 $\rho=$ density            & [$\mathrm{kg/m^3}]$          &  $833$   &  $961$    &  $913$  & $970$          \\
      $\gamma_{lg}=$ surface tension  & $[\mathrm{N/m}]$   &  $0.028$ &  $0.039$  &  $0.02$ & $0.0212$          \\
      $\nu=$ kinematics viscosity & $[\mathrm{m^2/s}]$  &  $3.67 \times 10^{-5}$  &  $1.1\times 10^{-3}$     & $5 \times 10^{-6}$  & $5\times10^{-4}$          \\
      $B=$ Bond number            & $[1]$        & $0.29$   & $0.24$    & $0.45$  & $0.45$  \\
      $\theta_0=$ contact angle    &   &  &   &  &  \\
      with stainless steel    & [Degree]   & 20 & 45  & 2 & 20  \\
      $\theta_0=$ contact angle    &   &  &   &  &  \\
      with copper   & [Degree]   & 20 & 45  & 2 & 25  \\
    
		\bottomrule\\
	\end{tabular}
\end{table}	

       \begin{figure}
         \centering
            \includegraphics[height=3.3 cm]{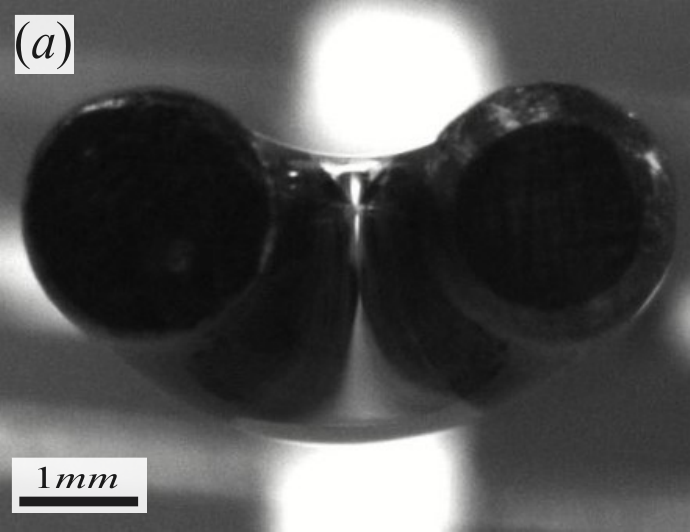}\hspace{0.1cm}
            \includegraphics[height=3.3 cm]{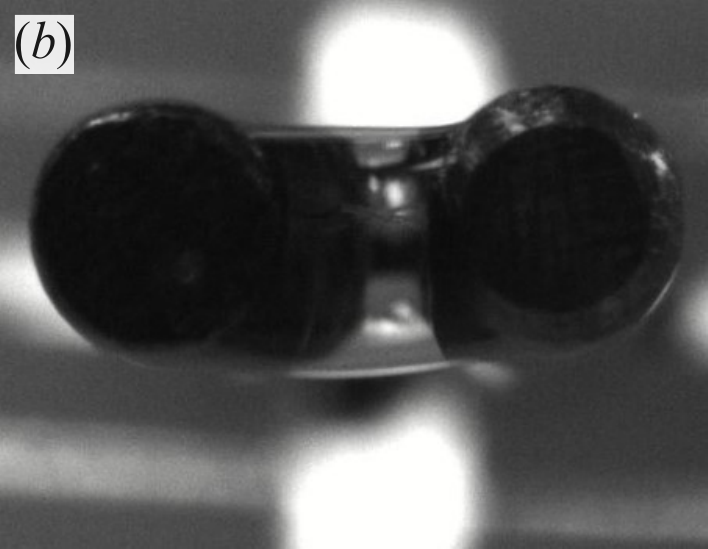}\hspace{0.1cm}
            \includegraphics[height=3.3 cm]{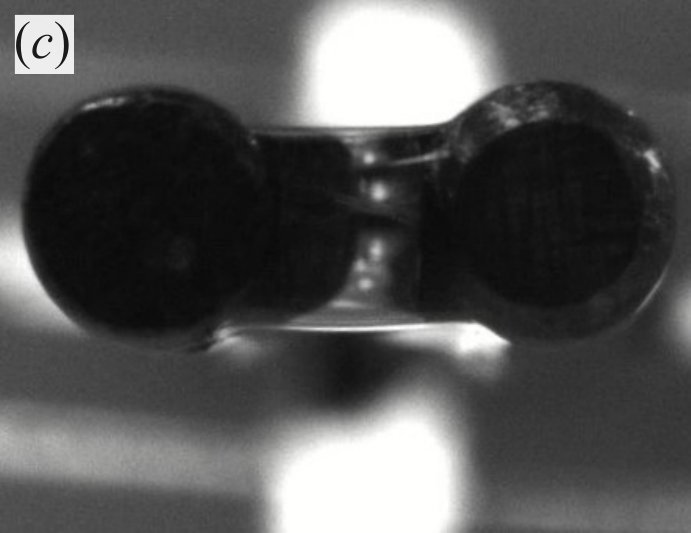}
            \\[0.3cm]
            \includegraphics[width=4.3 cm]{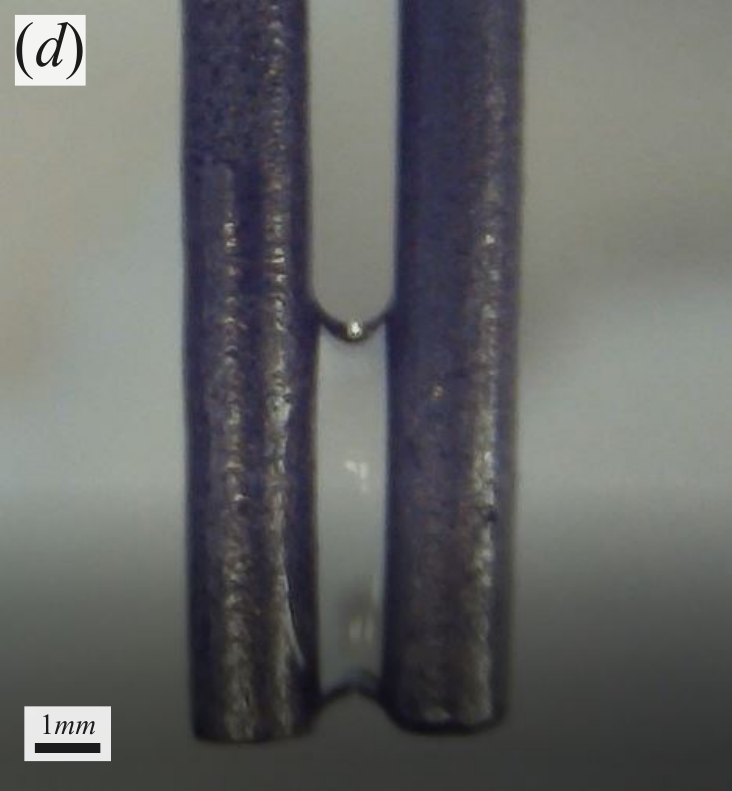}\hspace{0.1cm}
            \includegraphics[width=4.3 cm]{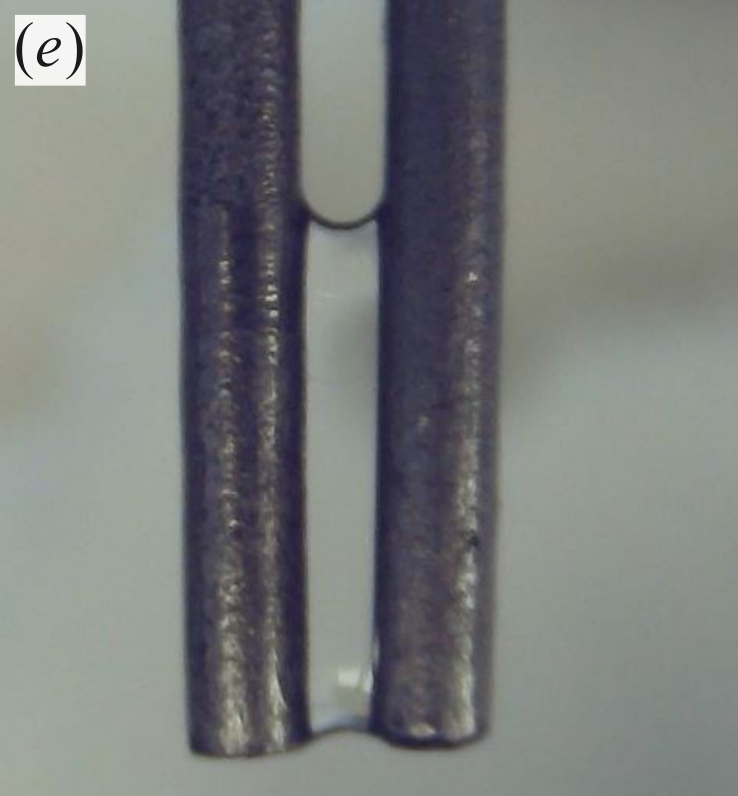}\hspace{0.1cm}
            \includegraphics[width=4.3 cm]{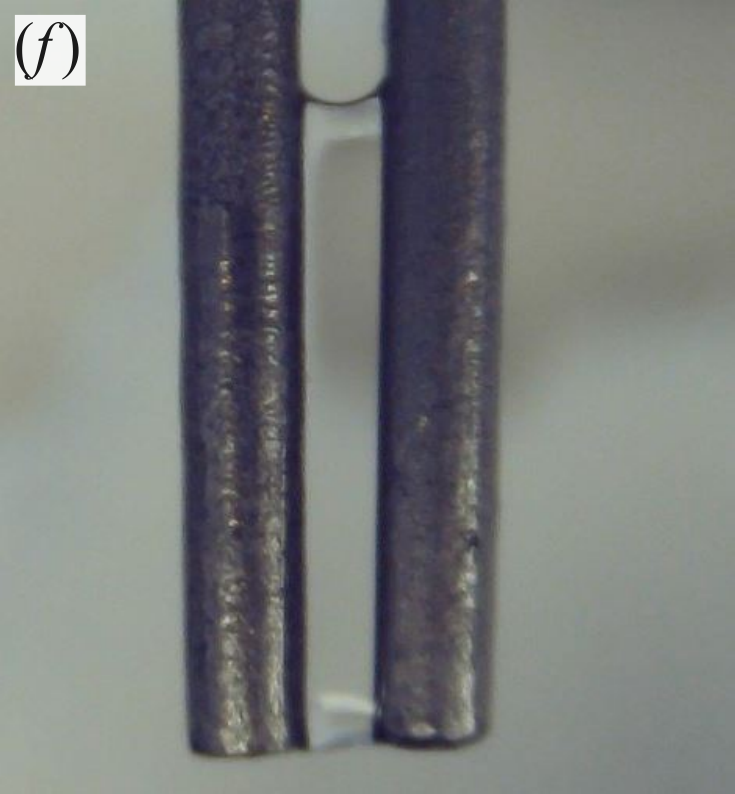}
          \caption{Front view of silicone-oil bridge suspended between platinised  titanium rods, for (\textit{a}) non-electrified case, (\textit{b}) $2\,000\, \mathrm{V}$, and (\textit{c}) $3\,000\, \mathrm{V}$. Top view of silicone-oil bridge suspended between platinised titanium rods corresponding to (\textit{d}) non-electrified case, (\textit{e}) $2\,000\, \mathrm{V}$, and (\textit{f}) $3\,000\, \mathrm{V}$. The video corresponding to this is available as supplementary material.}
          \label{silicone}
    \end{figure}

    \begin{figure}
         \centering
            \includegraphics[width=6.2 cm]{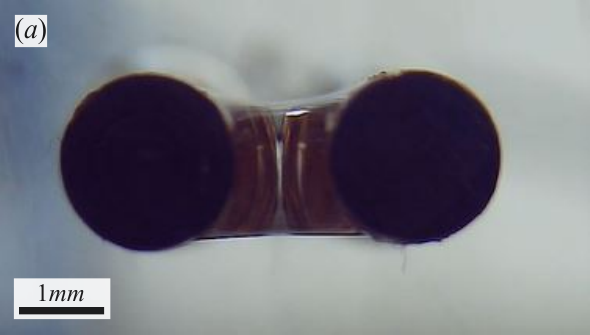}
            \includegraphics[width=6.2 cm]{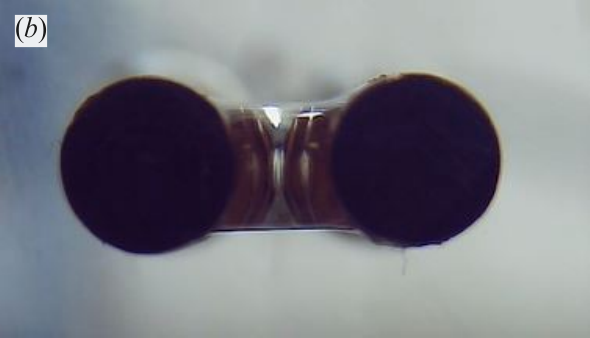}
            \includegraphics[width=6.2 cm]{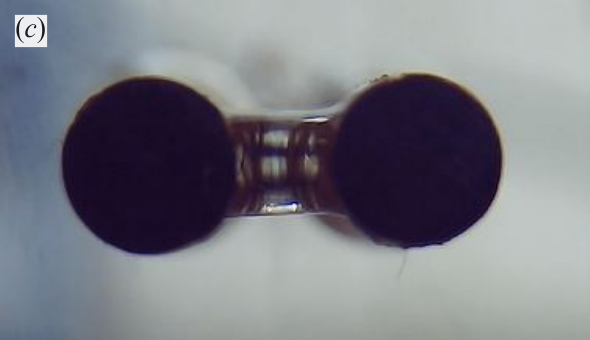}
            \includegraphics[width=6.2 cm]{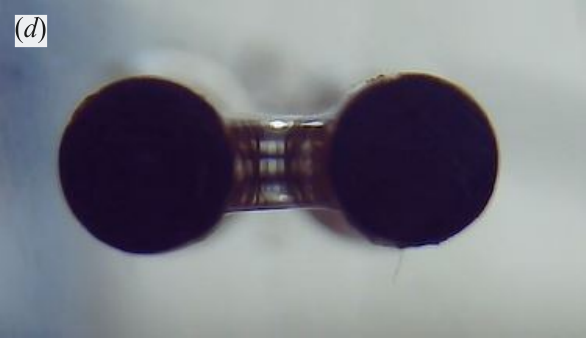}
          \caption{Front view of castor oil suspended between copper rods corresponding to (\textit{a}) non-electrified case, (\textit{b}) $2\,000\, \mathrm{V}$, (\textit{c})  $3\,000\, \mathrm{V}$, and (\textit{c}) $4\,000\, \mathrm{V}$. The video for this is available as supplementary material.}
          \label{castor_front}
          \vspace{0.4cm}

         \centering
            \includegraphics[width=3.25 cm]{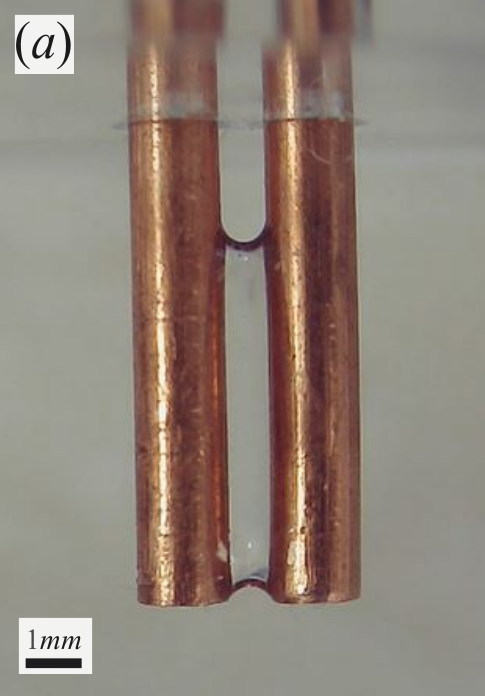}
            \includegraphics[width=3.25 cm]{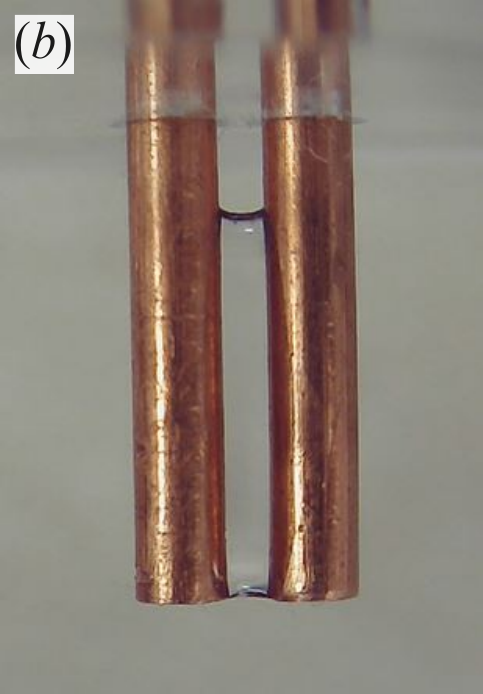}
            \includegraphics[width=3.25 cm]{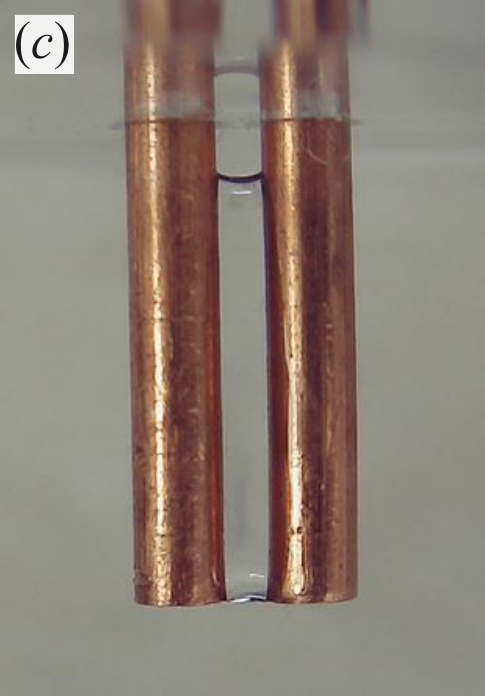}
            \includegraphics[width=3.25 cm]{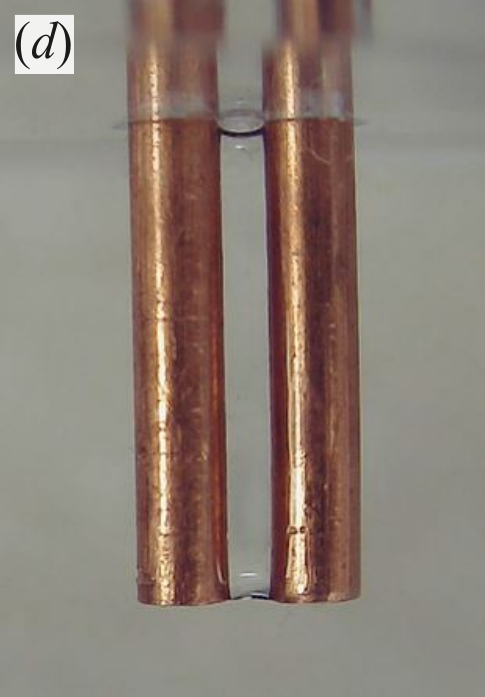}
          \caption{Top view of castor oil suspended between copper rods corresponding to (\textit{a}) non-electrified case, (\textit{b})  $2\,000\, \mathrm{V}$, (\textit{c})  $3\,000\, \mathrm{V}$, and (\textit{c}) $4\,000\, \mathrm{V}$. The video for this is available as supplementary material.}
          \label{castor_top}
    \end{figure}
    
Experiments were performed using pure non-conducting oils, namely mineral, castor and silicon oils. In table~\ref{tab:Para_Oils}, we present a summary of the relevant material properties of the oils. 
Note that the contact angles were measured using a drop shape analyser (DSA100E, KR\"USS Scientific), and the values reported in table~\ref{tab:Para_Oils} correspond to clean flat surfaces. 
However, we also observed significant variations due to surface roughness, and, therefore, these angles should be viewed as indicative and with a precision of around $\pm5^\circ$. 
Both electrified and non-electrified liquid bridges were recorded using front and top view cameras.

Before each experiment, the electrodes were cleaned with acetone, then distilled water and dried using compressed air.
A liquid bridge was formed by depositing an oil droplet of volume $\mathtt{\approx} 20 \, \mathrm{\mu L}$  between the electrodes using a pipette (RS PRO Pipette PE $5$ mL).
The bridge was in contact with the electrodes and surrounded by air, without contacting the supporting plate, as shown in figures~\ref{silicone}(\textit{a})--(\textit{c}).
After the settling of the bridge, video recording was started at zero applied voltage, followed by gradual increase of the voltage difference up to $4 000\, \mathrm{V}$, over a period of $5$ seconds. 
During the following $5$ seconds the maximum  voltage was held. Finally, the voltage difference was reduced gradually to zero over a further period of $5$ seconds, and then, after a few more seconds, the recording was ended.


We observed similar behaviour for the three oils. Representative results for silicone oil are shown in figure~\ref{silicone} and for castor oil in figures~\ref{castor_front} and \ref{castor_top}.  In the absence of an electric field, the liquid bridge 
quickly settles into an equilibrium shape, which hangs low due to the gravitational pull on the liquid. 
Since the capillary length of the liquid bridge is smaller than the separation between the electrodes, gravitational effects cannot be neglected. 
The front and top views of the initial non-electrified state are shown for the silicone-oil bridge in figures~\ref{silicone}(\textit{a}) and \ref{silicone}(\textit{d}) and for the castor-oil bridge in figures~\ref{castor_front}(\textit{a}) and \ref{castor_top}(\textit{a}). 

When an electric field is applied, the bridge moves upwards and the liquid--air interfaces become noticeably flatter.  
At the same time, the bridge extends along the electrodes (as seen in the top views), causing the reduction in the cross-sectional area. At higher voltages, the bridge reaches the maximal length and the minimal cross-sectional area (see figures~\ref{silicone}\textit{c} and  \ref{castor_top}\textit{d}, for silicone and castor oils, respectively). 
One can also observe that oils with higher viscosity react slower to the electric field changes, as expected.
    
These observations can be understood by noting that the applied electric field induces stresses at the liquid--air interfaces. These stresses compete with gravity and surface tension, leading to a new equilibrium shape, in which the liquid is drawn into the region between the electrodes, where the electric field is strongest.

\section{Full governing equations}
\label{sec:3}

To describe the experiments via a mathematical model, we account for the following physical effects beyond electrostatics and gravity: capillarity (equilibrium contact angle and surface tension), fluid flow and viscosity. 

   	\begin{figure}
    	\centering
    	\includegraphics[width=0.75\textwidth]{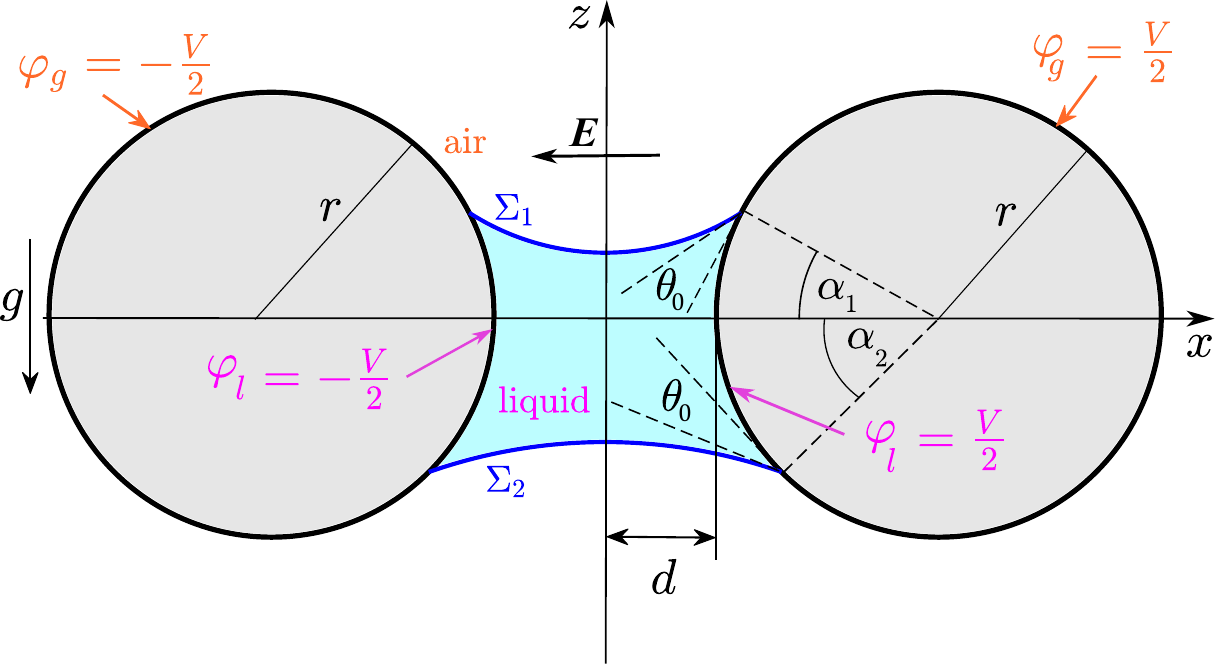}
	\vspace{-0.3cm}
    	\caption{2D sketch of the model geometry for an electrified pendant liquid bridge between two identical parallel cylinders of cross sectional radius $r$, with the distance $2d$ between the surfaces of the cylinders. The cylinders are electrodes with potential difference $V$ applied between them. The electric potential is denoted by $\varphi_l$ in the liquid and by $\varphi_g$ in the air. The equilibrium three-phase contact angle is $\theta_0$ and angles $\alpha_1$, $\alpha_2$ specify the locations of the three-phase contact lines. $\Sigma_1$ and $\Sigma_2$ denote the upper and lower liquid--air interfaces, respectively.}
  		\label{asymmetrics_figure_electric_11}
  	\end{figure}
We consider a liquid bridge between two parallel horizontal cylinders with  cross-sectional radius $r$. The distance between the surfaces of the cylinders is $2d$.
The cylinders are electrodes and there is potential difference $V$ applied between them.
The liquid is assumed to be a dielectric with permittivity $\epsilon_l$ (certainly true for the transformer oils used in our experiments), and we assume that it has constant density $\rho$ and dynamic viscosity $\mu$.
The permittivity of air is denoted by $\epsilon_g$.
We introduce Cartesian coordinate axes $(x,y,z)$, with the $x$-axis being horizontal and perpendicular to the axes of the cylinders, which lie in the $y$-direction; the $y$-axis is located right between them. The $z$-axis points upwards, so the acceleration due to gravity $\bm{g}$ points in the negative $z$ direction. We consider a 2D formulation, assuming invariance along the cylinder axis (i.e.\ the $y$-direction). This corresponds to sufficiently long bridges, where the cross-section is approximately uniform (in the regions away from the end menisci), as in the classical analysis of \cite{PRINCEN}. 
A schematic representation is shown in figure~\ref{asymmetrics_figure_electric_11}.

We now present the full governing equations, which form the basis for later derivations. The flow in the liquid is governed by the incompressible Navier--Stokes equations
\begin{equation}
    \label{stokes}
     \rho\frac{D\bm{u}}{Dt} = - \nabla  p + \mu \nabla^2 \bm{u} + \rho \bm{g}, \\
\end{equation}
\vspace{-0.5cm}
\begin{equation}
    \label{cont}
   \nabla  \cdot \bm{u} = 0,
\end{equation}
where $\bm{u}=(u,w)$ and $p$ are the velocity and the pressure in the liquid, respectively, and $D/Dt$ denotes the material derivative. Without loss of generality, it can be assumed that the pressure in the air is zero.
Note that for perfect dielectric liquids of constant permittivity there is no electric field contribution in the Navier--Stokes equations, and the coupling between the hydrodynamics and electrostatic phenomena takes place at the liquid--air interfaces, as explained below.

On the parts of the cylinder in contact with the liquid, we assume Navier slip conditions to allow for the motions of the contact lines, and we denote the equilibrium contact angle that the liquid makes with the cylinder surface by $\theta_0$.

As usual, we treat the liquid--air interfaces as surfaces of zero thickness. The upper and lower interfaces are denoted by $\Sigma_1$ and $\Sigma_2$ and are given by the equations $f_1(x,z,t)=0$ and $f_2(x,z,t)=0$, respectively. In the simplest case, $\Sigma_1$ and $\Sigma_2$ are given by the functions $z=h_1(x,t)$ and $z=h_2(x,t)$, respectively, and in such a case $f_1(x,z,t)=z-h_1(x,t)$ and $f_2(x,z,t)=z-h_2(x,t)$. 

At the liquid--air interfaces, we impose the kinematic boundary conditions:
\begin{equation}
    \frac{Df_i}{Dt}=0\quad\text{on}\quad \Sigma_i,\quad i=1,2,
\end{equation}
and the stress balance conditions:
\begin{equation}
    \bm{n}_i \cdot \bm{T}_l - \bm{n}_i \cdot \bm{T}_g= \gamma_{lg}\, \kappa_i \, \bm{n}_i\quad\text{on}\quad \Sigma_i,\quad i=1,2.
\end{equation}
Here, $\gamma_{lg}$ is the liquid--air surface tension, $\bm{n}_i$ is the unit normal vector to the interface $\Sigma_i$ pointing into the liquid, and $\kappa_i = \nabla \cdot \bm{n}_i$ is the curvature of $\Sigma_i$,  $i=1,2$. 
Also, $\bm{T}_l$ and $\bm{T}_g$ are the stress tensors in the liquid and in the air, respectively, given by 
\begin{equation}
     \bm{T}_l= \bm{\sigma}_l+ \bm{\sigma}_l^M, \qquad \bm{T}_g= \bm{\sigma}_g^M,
\end{equation}
where  $\bm{\sigma}_l=-p \bm{I} + \mu (\nabla \bm{u}+\nabla \bm{u}^T)$ is the Newtonian stress tensor in the liquid, 
$\bm{I}$ is the identity matrix, 
and $\bm{\sigma}_l^M$ and $\bm{\sigma}_g^M$ are the Maxwell stress tensors \citep{Castellanos, Saville} in the liquid and in the air, respectively, given by
\begin{eqnarray}
    \label{Mst1}
    \bm{\sigma}_l^M &=& \epsilon_l\, \bm{E}_{l} \, \bm{E}_{l}-\frac{1}{2} \, \epsilon_l (\bm{E}_{l} \cdot \bm{E}_{l} ) \bm{I},\\
    \label{Mst2}
    \bm{\sigma}_g^M &=&\epsilon_g \, \bm{E}_{g} \, \bm{E}_{g}-\frac{1}{2}\,  \epsilon_g   (\bm{E}_{g} \cdot \bm{E}_{g}) \bm{I}.
\end{eqnarray}
Here, 
$\bm{E}_{l}=-\nabla\varphi_l$ and $\bm{E}_{g}=-\nabla\varphi_g$ are the electric fields in the liquid and in the air, respectively, with $\varphi_l$ and $\varphi_g$ denoting the respective electric potentials. 

The electric potentials must satisfy Laplace's equation. Therefore
\refstepcounter{equation}
$$
    \nabla^2 \varphi_l=0, \qquad \nabla^2 \varphi_g=0. 
    \eqno{(\theequation{\mathit{a},\mathit{b}})}  \label{Laplaces_eqns}
$$

\noindent Without loss of generality, we assume that 
\begin{eqnarray}
    \varphi_l &=& -\frac{V}{2}, \quad \, \varphi_g=-\frac{V}{2}       \,    \quad \text{on the left cylinder}, \\
    \varphi_l&=&\frac{V}{2}, \quad \quad \varphi_g=\frac{V}{2} \quad    \quad \text{on the right cylinder}.
\end{eqnarray}
Also, in the far field $ \varphi_g \rightarrow 0$ as $(x^2+z^2)\rightarrow \infty$.

At the liquid--air interfaces the electric potential must be continuous, i.e. $\varphi_l=\varphi_g$,  
and we must have continuity of the normal component of the electric displacement
\begin{equation}  
\label{normalvec}
    \bm{n}_i \cdot \left(\epsilon_l \nabla \varphi_l\right)= 
    \bm{n}_i \cdot \left(\epsilon_g \nabla \varphi_g\right)\quad \text{on}\quad \Sigma_i,\quad i=1,2.
\end{equation}

In the rest of the manuscript, we exploit the symmetry about the $z$-axis. Therefore, it is enough to consider only the domain where $x\geq0$, imposing appropriate symmetry conditions at $x=0$ (namely, zero normal velocity, zero normal derivatives of the tangential velocity and pressure, zero electric potential, and zero slopes of the top and bottom liquid--air interfaces).

We note that the important dimensionless parameters that determine the equilibrium states of the system are the Bond number (measuring the relative importance of gravitational force to surface tension) and the electric Bond number (measuring the relative importance of electric force to surface tension) given by \citep[see e.g.][]{berthier2012physics}
\refstepcounter{equation}
$$
    B = \frac{\rho\, g\, r^2}{\gamma_{lg}}, \qquad 
    B_e = \frac{\epsilon_l V^2}{r \, \gamma_{lg}},
    \eqno{(\theequation{\mathit{a},\mathit{b}})}
$$

\noindent respectively, as well as the permittivity ratio, $\tilde{\epsilon}_l=\epsilon_l/\epsilon_g$, the dimensionless half-distance between the cylinders, $\tilde{d}=d/r$, and the equilibrium three-phase contact angle $\theta_0$. 

For the dynamics, another important dimensionless parameter is the Reynolds number, ${Re={\rho\, U\, r}/{\mu}}$, measuring the relative importance of inertia to viscosity, where $U$ is a characteristic velocity scale. However, for the reduced-order model derived later, we restrict attention to the overdamped regime in which inertia is negligible.

\section{Steady-state solutions}
\label{sec:4}

In the static case, we find that the shapes of the interfaces $\Sigma_1$ and $\Sigma_2$ are determined by the YL equations with additional contributions due to the electric field.
With the interfaces given by the functions $h_1(x)$ and $h_2(x)$, these equations have the following forms: 
\begin{equation}
    \label{Y_Laplace1}
   \rho g h_1 - p_0= \gamma_{lg} \frac{h_{1}''}{\left[1+(h_{1}')^2\right]^{3/2}} +\mathcal{E}_1,
\end{equation}
\begin{equation}
    \label{Y_Laplace2}
   \rho g h_2 - p_0=-\gamma_{lg} \frac{h_{2}''}{\left[1+(h_{2}')^2\right]^{3/2}} +\mathcal{E}_2.
\end{equation}
Here, $p_0$ is a constant representing the hydrostatic pressure in the liquid at $z=0$ which needs to be determined as part of the solution. Also, $\mathcal{E}_i$, $i=1,2$, are the electric contributions resulting from the jumps in the Maxwell stress tensors \eqref{Mst1} and \eqref{Mst2}, given by 
\begin{equation}
    \label{ee_1}
\mathcal{E}_i  =  \frac{1}{2}\left(\frac{\epsilon_l^2}{\epsilon_g}-\epsilon_l \right) \bm{E}_{ln}^2-\frac{1}{2} \left(\epsilon_g -\epsilon_l\right) \bm{E}_{lt}^2, \qquad i=1,2,
\end{equation}
where $\bm{E}_{ln}$ and $\bm{E}_{lt}$ are the normal and tangential component of $\bm{E}_l$ at the liquid--air interfaces. In particular, $\bm{E}_{ln} = -\bm{n}_1 \cdot  \nabla \varphi_l$ and $\bm{E}_{lt} = -\bm{t}_1 \cdot  \nabla \varphi_l$ at the upper interface $\Sigma_1$, and $\bm{E}_{ln} = -\bm{n}_2 \cdot  \nabla \varphi_l$ and $\bm{E}_{lt} = -\bm{t}_2 \cdot  \nabla \varphi_l$ at the lower interface $\Sigma_2$. Here, we remind that $\bm{n}_i$ is the unit normal vector to $\Sigma_i$, $i=1,2$, pointing into the liquid, and also $\bm{t}_i$ denotes a unit tangent vector to $\Sigma_i$, $i=1,2$.

If the interfaces $\Sigma_1$ and $\Sigma_2$ are given by implicit functions, we can use the following  parametrisations:
\begin{equation}
 \label{paraab}
   (x,z)=(x_1(s_1),z_1(s_1)) \quad \mbox{and\ }\quad  (x,z)=(x_2(s_2),z_2(s_2)),  
\end{equation}
where $s_1 \in [0,l_1]$ and  $s_2 \in [0,l_2]$ are the arc-length parameters along $\Sigma_1$ and $\Sigma_2$, respectively, defined so that $s_1=0$ and $s_2=0$ at $x=0$ and
$$
    (x_1')^2+(z_1')^2=1, \qquad 
    (x_2')^2+(z_2')^2=1.
\eqno{(\theequation{\mathit{a},\mathit{b}})}
\label{eq:arclength_parametrisations}
$$

\noindent The quantities $l_1$ and $l_2$ denote the half-lengths of $\Sigma_1$ and $\Sigma_2$, respectively, and need to be determined as part of the solution. 

The equations for the interfacial shapes become
\begin{eqnarray}
    \label{Y_Laplace11}
   \rho g z_1 - p_0 &=& \gamma_{lg} \frac{x_1'z_1''-z_1'x_1''}{\left(x_1'^2+z_1'^2\right)^{3/2} }+\mathcal{E}_1,\\
    \label{Y_Laplace22}
   \rho g z_2 - p_0 &=& - \gamma_{lg} \frac{x_2'z_2''-z_2'x_2''}{\left(x_2'^2+z_2'^2\right)^{3/2} }+\mathcal{E}_2.
\end{eqnarray}
In addition to the conditions for the electric field, the following boundary conditions must be satisfied for $x_1$, $z_1$, $x_2$ and $z_2$:
\begin{flushright}
\begin{tabular}{l l r}
$x_1(0)=0$, \hspace{0.5cm}   &
$x_2(0)=0$, \hspace{0.5cm}   & 
\refstepcounter{equation}(\theequation\textit{a,b})\label{BCAB_orig}\\[0.2cm]
$z_1'(0)=0$, \hspace{0.5cm}   &
$z_2'(0)=0$, \hspace{0.5cm}   & 
\refstepcounter{equation}(\theequation\textit{a,b})\label{BCCD_orig}\\[0.2cm]
$x_1(l_1)=d+r\left(1-\cos{\alpha_1} \right)$, \hspace{0.5cm}   &
$x_2(l_2)=d+r\left(1-\cos{\alpha_2} \right)$, \hspace{0.5cm}   & 
\refstepcounter{equation}(\theequation\textit{a,b})\label{BCE_orig}\\[0.2cm]
$z_1(l_1) = r \sin{\alpha_1}$, \hspace{0.5cm}   &
$z_2(l_2) = -r \sin{\alpha_2}$, \hspace{0.5cm}   & 
\refstepcounter{equation}(\theequation\textit{a,b})\label{BCF_orig}\\[0.2cm]
$z_1'(l_1) = \cot{(\alpha_1+\theta_0)}x_1'(l_1)$, \hspace{0.5cm}   &
$z_2'(l_2) = -\cot{(\alpha_2+\theta_0)} x_2'(l_2)$. \hspace{0.5cm}   & 
\refstepcounter{equation}(\theequation\textit{a,b})\label{BCG_orig}\\[0.2cm]
\end{tabular}
\end{flushright}
Here, $\alpha_1$ and $\alpha_2$ are the angles to the horizontal that determine the points of contact between the right cylinder and the upper and lower interfaces, respectively -- see figure~\ref{asymmetrics_figure_electric_11}.

Additionally, we have the constraints for the arc lengths $l_1$ and $l_2$:
\refstepcounter{equation}
$$  \int_0^{l_1} \left[(x_1')^2+(z_1')^2\right]^{1/2} ds =l_1, \quad \int_0^{l_2} \left[(x_2')^2+(z_2')^2\right]^{1/2} ds =l_2,
     \eqno{(\theequation{\mathit{a},\mathit{b}})} \label{Arclengths_cond_orig}$$

\noindent and the constraint imposing that the area of the cross section must be equal to a given constant (say $A_0$):
\begin{equation}
2A_1+2A_2=A_0,
\label{area_cosnstraint_orig}
\end{equation}
where $A_1$ and $A_2$ are the areas of the upper and lower parts (above and below the $x$ axis), respectively, given by
\begin{eqnarray}
    A_1 &=& \int_0^{l_1}z_1(s)x_1'(s)\, ds-\left( \frac{r^2 \alpha_1}{2}-\frac{1}{2}r^2 \cos{\alpha_1}\sin{\alpha_1}\right),\label{areaA1_orig}\\
    A_2 &=& -\int_0^{l_2}z_2(s)x_2'(s)\, ds-\left( \frac{r^2 \alpha_2}{2}-\frac{1}{2}r^2 \cos{\alpha_2}\sin{\alpha_2}\right).\label{areaA2_orig}
\end{eqnarray}
The first term in each of the above integrals is the area between the respective interface and the $x$-axis, while the second term is the half-segment area of the cylinder directly below/above the respective interface, which must be subtracted. For the case when the interfaces are single valued, then the first terms in equations \eqref{areaA1_orig} and \eqref{areaA2_orig} are simply $\int_0^{x_1(l_1)}h_1\,dx$ and $\int_0^{x_2(l_2)}h_2\,dx$, respectively.

\subsection{Computation of steady-state solutions in the absence of an electric field}
\label{Sec:Steady_nonelectrified}
In the absence of the electric field, using that $(x_i')^2+(z_i')^2=1$ and $x_i' x_i''+z_i' z_i''=0$ for $i=1,2$ and introducing $c_0\equiv - p_0/\gamma_{lg}$,  
we can rewrite the equations for the interfacial shapes as a system of first-order ordinary differential equations for the following unknowns:
\begin{align}
      \bm{X}&=\left(X_1,X_2,X_3,X_4 \right)=\left(x_1,x_1',z_1,z_1' \right), \\
      \bm{Y}&=\left(Y_1,Y_2,Y_3,Y_4 \right)=\left(x_2,x_2',z_2,z_2' \right).
\end{align}
The system of equations takes the form\\[-0.2cm]
\begin{flushright}
\begin{tabular}{l l r}
$\displaystyle X_1' = X_2,$ \hspace{1.2cm} &
$\displaystyle Y_1' = Y_2,$ \hspace{1cm} &
\refstepcounter{equation} (\theequation\textit{a,b})\label{x1y1}\\[0.3cm]
$\displaystyle X_2' = -\frac{\rho g}{\gamma_{lg}} X_3 X_4 - c_0 X_4,$ \hspace{1.2cm} &
$\displaystyle Y_2' = -\frac{\rho g}{\gamma_{lg}} Y_3 Y_4 - c_0 Y_4,$ \hspace{1cm} &
\refstepcounter{equation}(\theequation\textit{a,b})\\[0.3cm]
$\displaystyle X_3' = X_4,$ \hspace{1.2cm} &  
$\displaystyle Y_3' = Y_4,$ \hspace{1cm} &
\refstepcounter{equation}(\theequation\textit{a,b})\\[0.3cm]
$\displaystyle X_4' =  \frac{\rho g}{\gamma_{lg}} X_2X_3 + c_0 X_2,$ \hspace{1.2cm} & 
$\displaystyle Y_4' =  \frac{\rho g}{\gamma_{lg}} Y_2Y_3 + c_0 Y_2$. \hspace{1cm} &  
\refstepcounter{equation}(\theequation\textit{a,b})\label{x4y4}\\[0.3cm]
\end{tabular}
\end{flushright}

The boundary conditions for the new system are\\[-0.2cm]
\begin{flushright}
\begin{tabular}{l l r}
$X_1(0)=0$, \hspace{0.5cm}   &
$Y_1(0)=0$, \hspace{0.5cm}   & 
\refstepcounter{equation}(\theequation\textit{a,b})\label{BCAB}\\[0.2cm]
$X_4(0)=0$, \hspace{0.5cm}   &
$Y_4(0)=0$, \hspace{0.5cm}   & 
\refstepcounter{equation}(\theequation\textit{a,b})\label{BCCD}\\[0.2cm]
$X_1(l_1)=d+r\left(1-\cos{\alpha_1} \right)$, \hspace{0.5cm}   &
$Y_1(l_2)=d+r\left(1-\cos{\alpha_2} \right)$, \hspace{0.5cm}   & 
\refstepcounter{equation}(\theequation\textit{a,b})\label{BCE}\\[0.2cm]
$X_3(l_1) = r \sin{\alpha_1}$, \hspace{0.5cm}   &
$Y_3(l_2) = -r \sin{\alpha_2}$, \hspace{0.5cm}   & 
\refstepcounter{equation}(\theequation\textit{a,b})\label{BCF}\\[0.2cm]
$X_4(l_1) = \cot{(\alpha_1+\theta_0)}X_2(l_1)$, \hspace{0.5cm}   &
$Y_4(l_2) = -\cot{(\alpha_2+\theta_0)} Y_2(l_2)$. \hspace{0.5cm}   & 
\refstepcounter{equation}(\theequation\textit{a,b})\label{BCG}\\[0.2cm]
\end{tabular}
\end{flushright}
The constraints for the arc lengths $l_1$ and $l_2$ take the following form:
\refstepcounter{equation}
$$  \int_0^{l_1} (X_2^2+X_4^2)^{1/2} ds =l_1, \quad \int_0^{l_2} (Y_2^2+Y_4^2)^{1/2} ds =l_2,
     \eqno{(\theequation{\mathit{a},\mathit{b}})} \label{Arclengths_cond}$$

\noindent We also have the area constraint (\ref{area_cosnstraint_orig}), where $A_1$ and $A_2$ take the following form: 
\begin{eqnarray}
    A_1 &=& \int_0^{l_1}X_2(s)X_3(s) ds-\left( \frac{r^2 \alpha_1}{2}-\frac{1}{2}r^2 \cos{\alpha_1}\sin{\alpha_1}\right),\label{areaA1}\\
    A_2 &=& -\int_0^{l_2}Y_2(s)Y_3(s) ds-\left( \frac{r^2 \alpha_2}{2}-\frac{1}{2}r^2 \cos{\alpha_2}\sin{\alpha_2}\right).\label{areaA2}
\end{eqnarray}
We note that \cite{Cooray2016} obtained analytical solutions of the YL equations in terms of elliptic integrals of the first and second kind to describe liquid bridges between horizontal cylinders.
Here, however, we obtain the shapes of liquid bridges by adopting the following numerical approach to solve problem \eqref{x1y1}--\eqref{Arclengths_cond}, \eqref{area_cosnstraint_orig}: We apply a Newton iterations procedure to determine $c_0$, $\alpha_1$, $\alpha_2$, $l_1$ and $l_2$.
Given initial estimates for these variables, we solve the boundary-value problem (\ref{x1y1}\textit{a})-(\ref{x4y4}\textit{a}), (\ref{BCAB}\textit{a})-(\ref{BCF}\textit{a}) for $\bm{X}$ and the boundary-value problem (\ref{x1y1}\textit{b})-(\ref{x4y4}\textit{b}), (\ref{BCAB}\textit{b})-(\ref{BCF}\textit{b}) for $\bm{Y}$. Then, a further Newton procedure step is applied to correct $c_0$, $\alpha_1$, $\alpha_2$, $l_1$ and $l_2$, in order to satisfy the conditions (\ref{BCG}\textit{a,b}),  (\ref{Arclengths_cond}\textit{a,b}) and \eqref{area_cosnstraint_orig}.
The iterations are continued until convergence is achieved. We compared with the analytical solutions of \cite{Cooray2016} in order to validate the accuracy of our numerical implementation.

\subsection{Computation of steady-state solutions in the presence of an electric field}
\label{sect:electrified_steady_states}

In the presence of the electric field, there appear additional non-local terms $\mathcal{E}_1$ and $\mathcal{E}_2$ in the modified YL (or electrified YL, referred to as eYL) equations \eqref{Y_Laplace11} and \eqref{Y_Laplace22} resulting from the jumps in the Maxwell stress tensor across the liquid--air interfaces. To compute these contributions, we need to solve Laplace's equations (\ref{Laplaces_eqns}\textit{a,b}) subject to appropriate boundary conditions. This can be done using a boundary-element method. First, using a standard methodology \citep[see e.g.][]{{POZRIKIDIS}}, we reformulate the problem for the electric field in the form of boundary-integral equations. Due to symmetry with respect to the $z$-axis, it is sufficient to consider only the right half of the $(x,z)$-plane, taking into account that the electric  potentials must vanish on the $z$-axis.  We denote the halves of the upper and lower liquid--air interfaces $\Sigma_1$ and $\Sigma_2$ belonging to the right-half of the $(x,z)$-plane by $\sigma_1$ and $\sigma_2$, respectively, and we denote the wetted and dry parts of the right cylinder by $c_1$ and $c_2$, respectively. 

The boundary-integral equation for the liquid region at a point $\bm{x}_0=(x_0,z_0)$ lying on $\sigma_1\cup \sigma_2\cup c_1$ takes the form  
\begin{eqnarray}
    \displaystyle\frac{1}{2} \varphi_l(\bm{x}_0)&=&\displaystyle\int_{\sigma_1\cup \sigma_2} \varphi_l(\bm{x}) 
    \left[\bm{n}(\bm{x}) \cdot \nabla \mathcal{G}\left(\bm{x},\bm{x}_0\right) \right] \; d l(\bm{x})+ \frac{V}{2} \int_{c_1} \bm{n}(\bm{x}) \cdot \nabla \mathcal{G}\left(\bm{x},\bm{x}_0\right)   \; d l(\bm{x})\nonumber \\
    &&+\displaystyle \int_{\sigma_1\cup \sigma_2\cup c_1} \mathcal{G}(\bm{x},\bm{x}_0)  \bm{E}_{ln}(\bm{x}) \; d l(\bm{x}),
    \label{BI1}
\end{eqnarray}
where $\bm{x}=(x,z)$ is a variable point on the respective boundaries, $\bm{n}(\bm{x})$ is the unit normal vector to the respective boundaries pointing into the liquid and $\bm{E}_{ln}=-\bm{n} \cdot \nabla \varphi_l$ is the normal component of the electric field in the liquid at the respective boundaries. Also, $\mathcal{G}\left(\bm{x},\bm{x}_0\right)$ is the Green's function for Laplace's equation that vanishes at $x=0$, and it is given by 
\begin{equation}
   \mathcal{G} (\bm{x},\bm{x}_0)=-\frac{1}{2\pi} \ln |\bm{x}-\bm{x}_0|+\frac{1}{2\pi} \ln |\bm{x}-\bm{x}_0^{im}|,
\end{equation}
where $\bm{x}_0^{im}=(-x_0,y_0)$ is the image of the point $\bm{x}_0$ with respect to the symmetry $z$-axis.

The boundary-integral equation for the air region at a point $\bm{x}_0=(x_0,z_0)$ lying on $\sigma_1\cup \sigma_2\cup c_2$ takes the form
\begin{eqnarray}
    \displaystyle\frac{1}{2}\varphi_g(\bm{x}_0)&=& -\displaystyle\int_{\sigma_1\cup \sigma_2} \varphi_g(\bm{x}) \left[{\bm{n}}(\bm{x}) \cdot \nabla \mathcal{G}(\bm{x},\bm{x}_0)\right] \; d l (\bm{x})
    +\frac{V}{2} \int_{c_2} \bm{n}(\bm{x}) \cdot \nabla \mathcal{G}(\bm{x},\bm{x}_0) \; dl (\bm{x})\nonumber \\
    &&  - \displaystyle\int_{\sigma_1 \cup \sigma_2} \mathcal{G}(\bm{x},\bm{x}_0) \bm{E}_{gn} \; d l(\bm{x}) 
    + \int_{c_2} \mathcal{G}(\bm{x},\bm{x}_0) \bm{E}_{gn}(\bm{x}) \;dl (\bm{x}), 
    \label{BI2}
\end{eqnarray}
where $\bm{n}(\bm{x})$ points into the liquid if $\bm{x}\in \sigma_1 \cup \sigma_2$, as before, and $\bm{n}(\bm{x})$ points into the air if $\bm{x}\in c_2$. Also, $\bm{E}_{gn}=-\bm{n} \cdot \nabla \varphi_g$ is the normal component of the electric field in the air at the respective boundaries. Note that to obtain these boundary-integral equations we took the advantage of the fact the electric potentials vanish on the $z$-axis and tend to zero at infinity, i.e. when $(x^2+y^2)\rightarrow \infty$, and we also used that $\mathcal{G} (\bm{x},\bm{x}_0)=0$ when $\bm{x}$ is on the $z$-axis.

Due to the boundary conditions, $\varphi_g$ and $\bm{E}_{gn}$ at $\sigma_1 \cup \sigma_2$ can be expressed in terms of $\varphi_l$ and $\bm{E}_{ln}$ as $\varphi_g=\varphi_l$ and $\bm{E}_{gn}=\tilde{\epsilon}_l \bm{E}_{ln}$. Hence, we are left with solving the boundary-integral equations for the following unknowns: the electric potential in the liquid, $\varphi_l$, at $\sigma_1 \cup \sigma_2$, the normal component of the electric field in the liquid, $\bm{E}_{ln}$, at $\sigma_1 \cup \sigma_2 \cup c_2$, and the normal component of the electric field in the air, $\bm{E}_{gn}$, at the cylinder $c_2$. Knowledge of the solution furnishes the electric-field contribution to the eYL equations \eqref{Y_Laplace11} and \eqref{Y_Laplace22}. In these equations, the tangential components of the  electric filed at the liquid--air interfaces, i.e. $\bm{E}_{lt}$ at $\sigma_1$ and $\sigma_2$, are needed. These can be computed by numerically differentiating the electric potential $\varphi_l$ on $\sigma_1$ and $\sigma_2$ along the tangential directions. 

More specifically, given guesses for the liquid--air interface shapes, the electric-field problem is solved numerically using a boundary-element method, in which $\sigma_1$, $\sigma_2$, $c_1$ and $c_2$ are  discretised into a fixed number of straight elements connected by a sequence of nodes. The unknown variables are approximated with constant values over each element. The integral equations \eqref{BI1} and \eqref{BI2} are enforced at the midpoints of the elements, with the integrals of $\mathcal{G}(\bm{x},\bm{x}_0)$, $\mathcal{G}_x(\bm{x},\bm{x}_0)$ and $\mathcal{G}_y(\bm{x},\bm{x}_0)$ computed using a Gauss-Legendre quadrature. This leads to a system of linear equations for the values of $\varphi_l$ and $\bm{E}_{lt}$ on elements of the top and bottom liquid--air interfaces, and the values of $\bm{E}_{lt}$ and $\bm{E}_{gt}$ on the elements of the wetted and dry cylinder parts, respectively, which is solved using Gaussian elimination.

The {\it a priori} unknown locations of the liquid--air interfaces $\sigma_1$ and $\sigma_2$ (along with the unknown lengths $l_1$ and $l_2$ of $\sigma_1$ and $\sigma_2$, respectively, and the unknown pressure $p_0$) can then be computed numerically by solving a system of nonlinear equations resulting from a discretisation of the eYL equations \eqref{Y_Laplace11} and \eqref{Y_Laplace22} using finite differences for the derivatives with respect to $s_1$ and $s_2$ and employing boundary conditions \eqref{BCAB_orig}--\eqref{BCG_orig} and discretised versions of the integral constraints \eqref{Arclengths_cond_orig} and \eqref{area_cosnstraint_orig}, and with the electric contributions computed using the boundary-element method described above. The nonlinear system is solved using a Newton iteration procedure, with the electric-field contributions recomputed at each iteration step using the boundary-element method.  Numerical convergence tests were performed and it was found that discretising $\sigma_1$ and $\sigma_2$ into $40$ elements each, and, using $40$ elements for the wetted part of the cylinder is sufficient to obtain numerically accurate solutions. The number of the elements for the dry cylinder part is chosen to be  $\lceil 40\, (2 \pi -\alpha_1-\alpha_2)/(\alpha_1+\alpha_2) \rceil$, so that the length of the element on the dry part is approximately the same as the length of the element on the wetted part.

\begin{table}
	\caption{Dimensionless parameters for the steady-state numerical simulations presented in figure~\ref{V_0_1}.} 
	\label{tab:Para1}
	\centering
	\begin{tabular}{l c r}
		\toprule
    \textbf{Parameter} &  \textbf{Symbol} & \textbf{Value}  \\ 
		\midrule
		  half-distance between the rods      & $\tilde{d}$          & $0.5$ \\
          equilibrium contact angle              & $\theta_0$   & $\pi/12$  \\
	      Bond number                         & $B$          & $0.5$    \\
	      electric Bond number              & $B_e$        & $12$    \\
          permittivity ratio of the liquid & $\tilde{\epsilon}_l$ & $3$      \\
		\bottomrule\\
	\end{tabular}
	\vspace{-0.3cm}
\end{table}	

\begin{figure}
    	\centering
        \includegraphics[height=5.8cm]{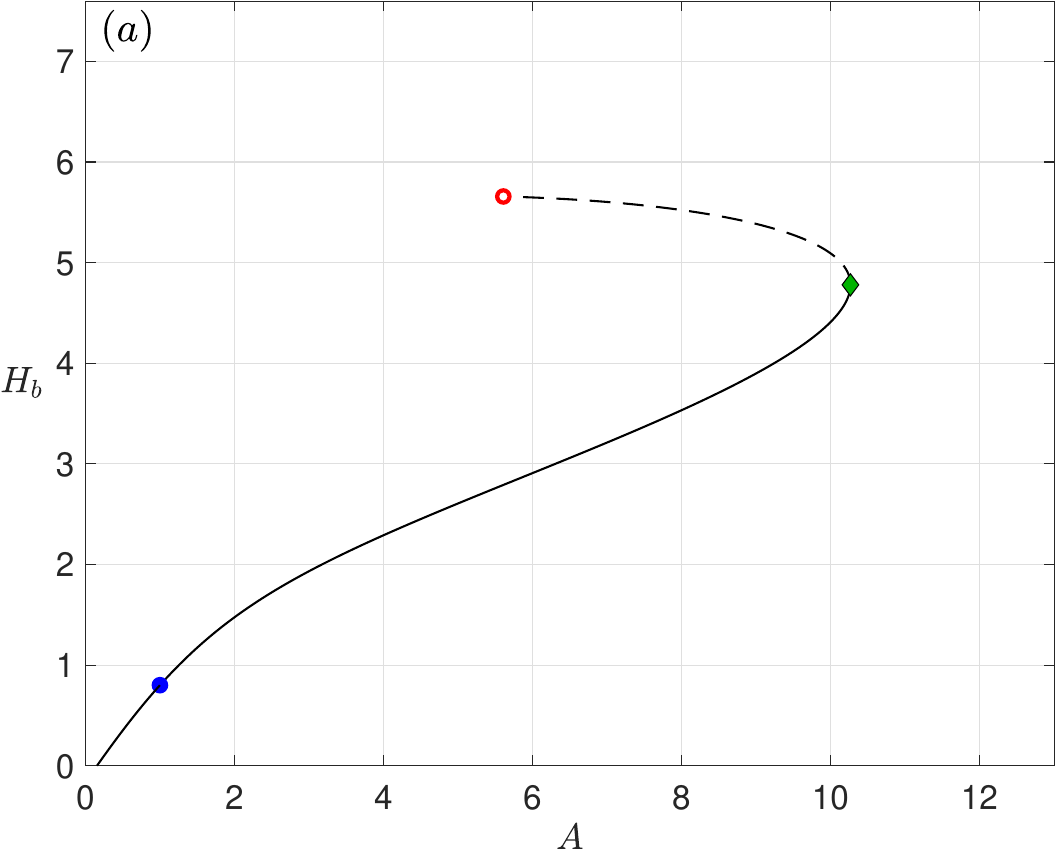}\hspace{0.3cm}
        \includegraphics[height=5.8cm]{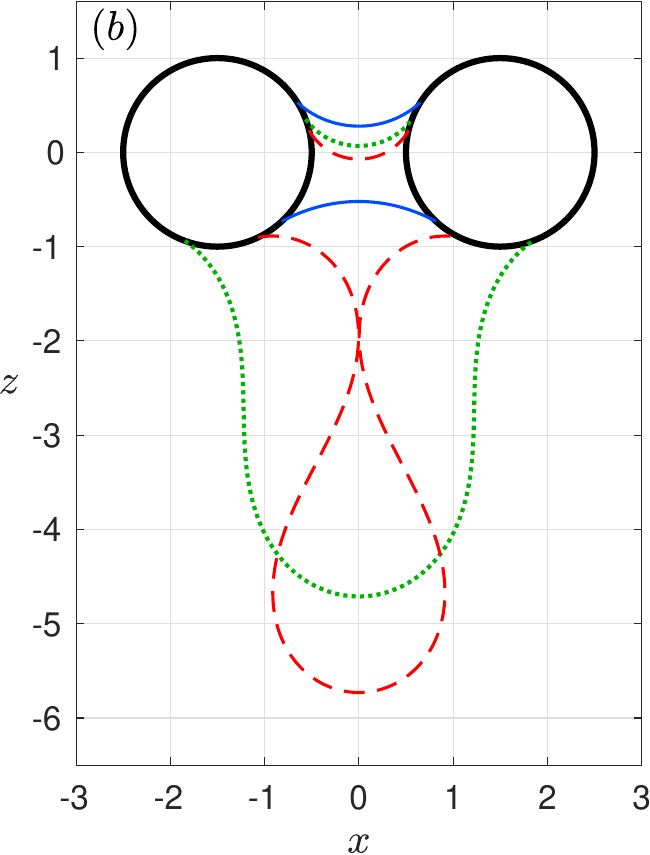}\\[0.3cm]
        \includegraphics[height=5.8cm]{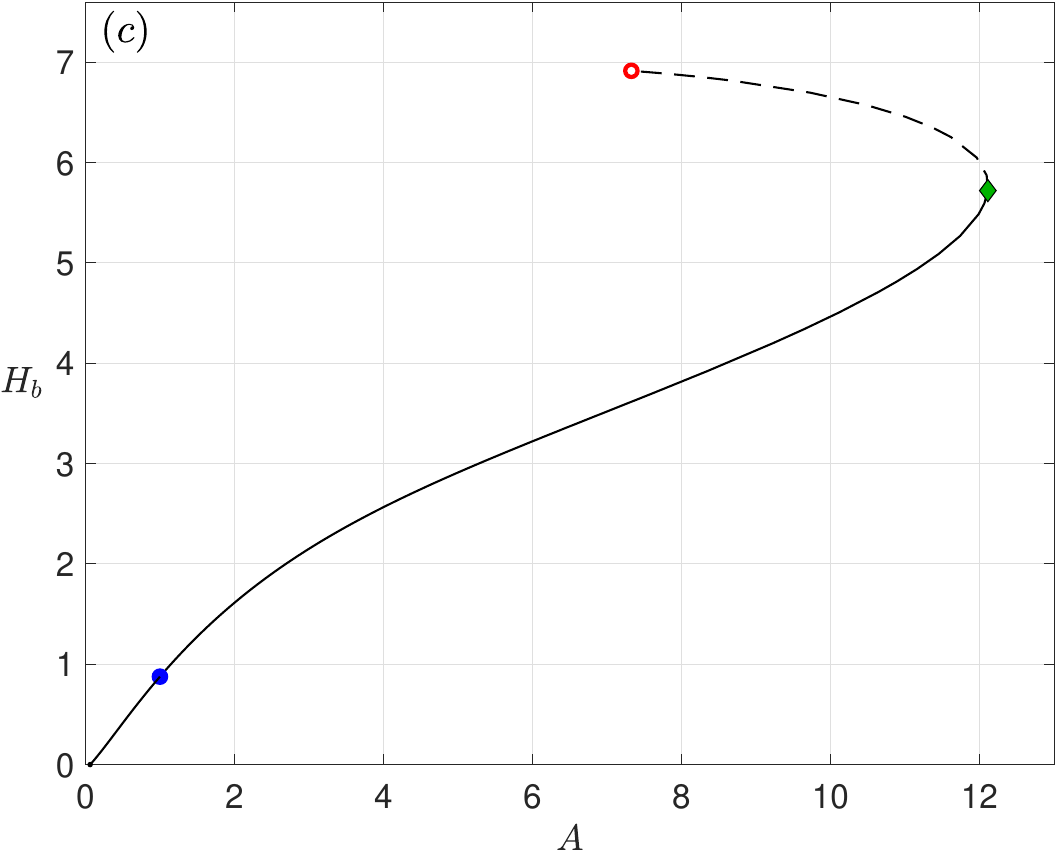}\hspace{0.3cm}
        \includegraphics[height=5.8cm]{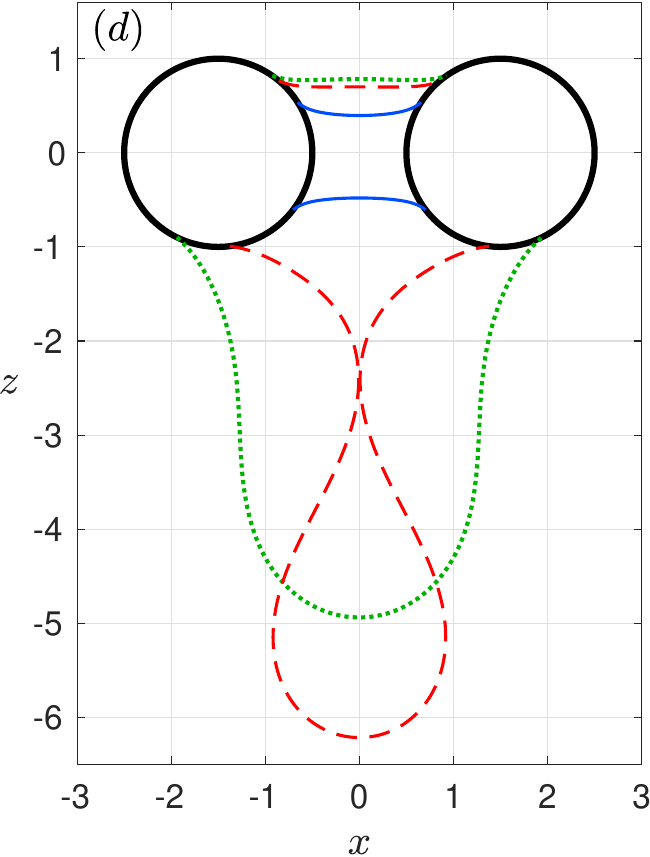}
    	\caption{Bifurcation diagrams (left panels) and solution profiles (right panels) for the non-electrified case with $B_e=0$ (top panels) and the electrified case with $B_e=12$ (bottom panels). See table~\ref{tab:Para1} for the values of all the dimensionless parameters. The bifurcation diagrams show the dependence of the bridge thickness in the middle, $H_b$, on the area of the bridge cross-section, $A$. The solid and dashed lines correspond to stable and unstable solutions, respectively. Solution profiles are shown for different values of $A$. The blue solid lines correspond to $A=1$ (see the blue filled circles in the bifurcation diagrams). The green dotted lines correspond to the turning points in the bifurcation diagrams (indicated by the green filled diamonds), for which $A=10.3$ and $A=12.1$ for the electrified and non-electrified case, respectively. The red dashed lines correspond to the points of pinching of the lower liquid--air interface (indicated by the red empty circles in the bifurcation diagrams), for which $A=5.6$ and $A=7.3$ for the electrified and non-electrified case, respectively.}
    \vspace{0.3cm}
  		\label{V_0_1}
\end{figure}

\subsection{Numerical steady-state results}
\label{sec:5}

We now present steady-state solutions for liquid bridges in both non-electrified and electrified configurations. In particular, we examine how the bridge shape varies with cross-sectional area.  To characterise the solutions, we use the dimensionless bridge thickness in the middle, $H_b=[z_1(0)-z_2(0)]/r$, and construct bifurcation diagrams (using the pseudo-arclength continuation method) showing how $H_b$ depends on the dimensionless cross-sectional area $A=A_0/r^2$. 
We compare the non-electrified and electrified cases in figure~\ref{V_0_1}. For these results, the dimensionless half-distance between the rods is $\tilde d =0.5$, the equilibrium contact angle is $\theta_0=\pi/12$ and the Bond number is $B=0.5$. Additionally, for the electrified case, we assume that the  permittivity ratio  is $\tilde{\epsilon}_l=3$  and that the electric Bond number is $B_e=12$;
see table~\ref{tab:Para1} for all the values of the dimensionless parameters. 

Panels (\textit{a}) and (\textit{c}) of figure~\ref{V_0_1} represent bifurcation diagrams showing the dependence of the bridge thickness $H_b$ on the area of the bridge cross-section $A$ for the electrified and non-electrified case, respectively. The black solid lines indicate stable solutions up to the turning points (represented by the green filled diamonds). For these points $A=10.3$ and $A=12.1$, for the non-electrified and electrified case, respectively. These points correspond to the maximal cross-sectional bridge areas \citep[or to the points of the maximum trapping capacity, as defined by][]{Cooray2016} for the respective parameter values. The black dashed lines indicate unstable solutions. They start at the points corresponding to the maximal cross-sectional bridge areas and continue backwards up to the points where the solution profile for the lower liquid--air interface pinches. These points are represented by the red empty circles. For these points $A=5.6$ and $A=7.3$, for the non-electrified and electrified case, respectively. The branches of unstable solutions can be continued beyond these point to smaller $A$ values, but the profiles become self-intersecting and, therefore, not physically relevant. Thus, we terminate the bifurcation diagrams at these points.

Panels (\textit{b}) and (\textit{d}) of figure~\ref{V_0_1} show some solution profiles of the liquid bridge for the electrified and non-electrified case, respectively. The blue solid lines correspond to $A=1$. The respective points in the bifurcation diagrams are indicated by the blue filled circles. The green dotted lines correspond to the maximal cross-sectional bridge areas (see the green filled diamonds in the bifurcation diagrams). Finally, the red dashed lines correspond to the bridges at the point of pinching. As has been already mentioned, the respective points in the bifurcation diagrams are indicated by the red empty circles.

The results indicate that electrified liquid-bridges can hold more liquid, i.e.\ the maximum trapping capacity is increased in the presence of an electric field. Also, it can be observed that for relatively small bridge cross-sectional areas, the effect of the electric field is to pull the liquid bridge upwards and to make the liquid--air interfaces flatter (compare, for example, the blue solid lines for $A=1$ in panels (b) and (d) of figure~\ref{V_0_1}). 

\subsection{Comparison with experiments}
\label{sec9}

    \begin{figure}
         \centering
           \includegraphics[width=0.45\textwidth]{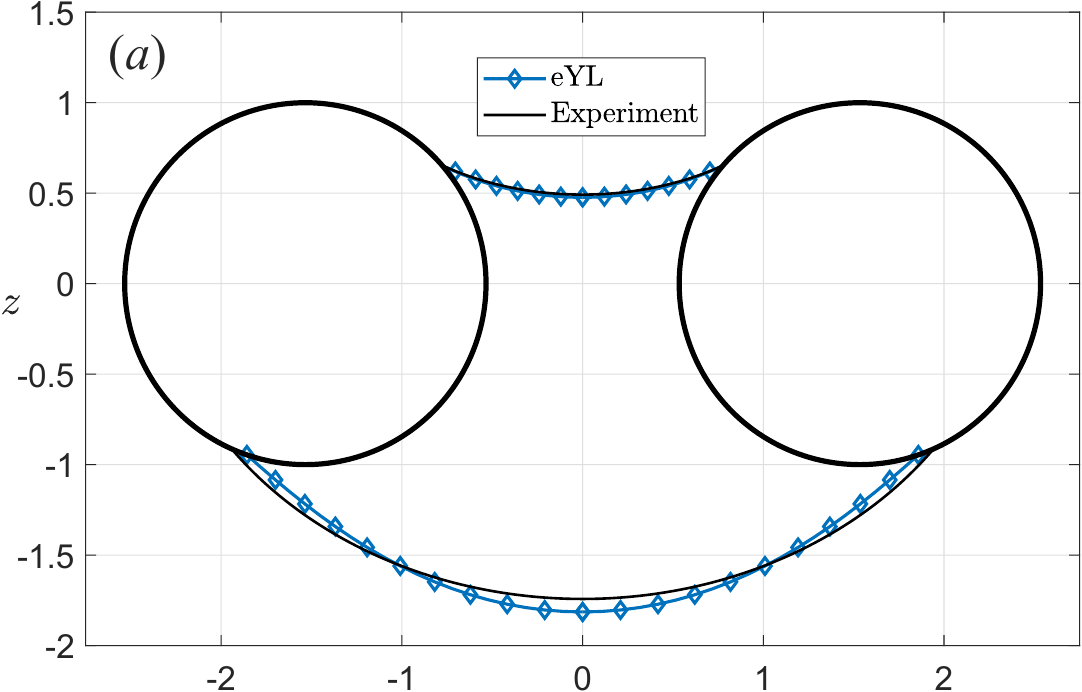}
           \includegraphics[width=0.45\textwidth]{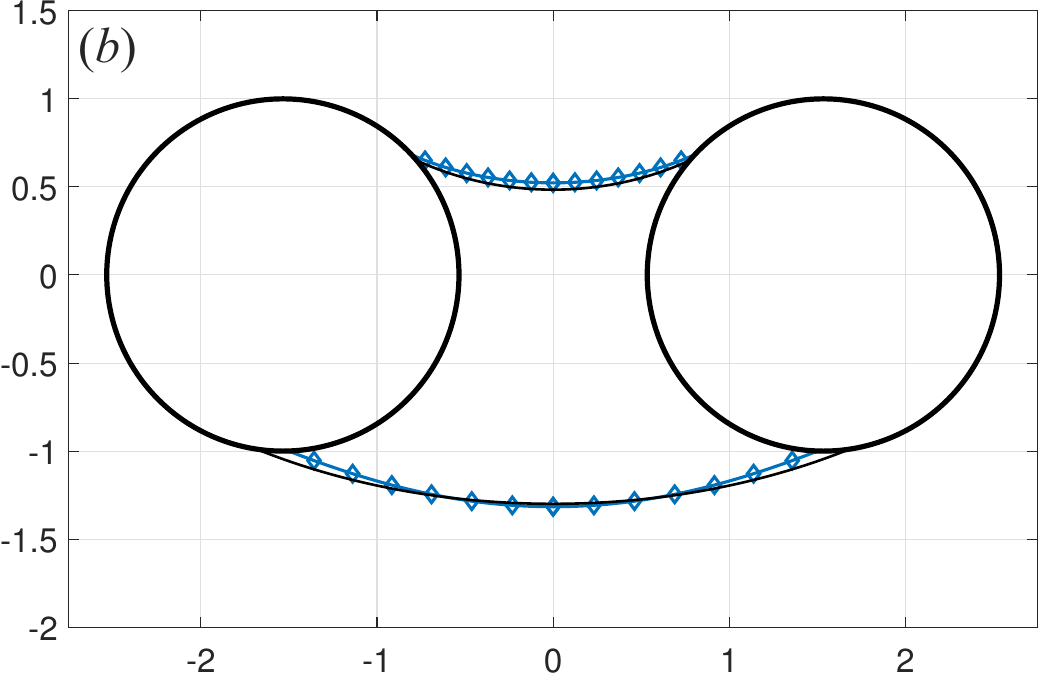}
           \includegraphics[width=0.45\textwidth]{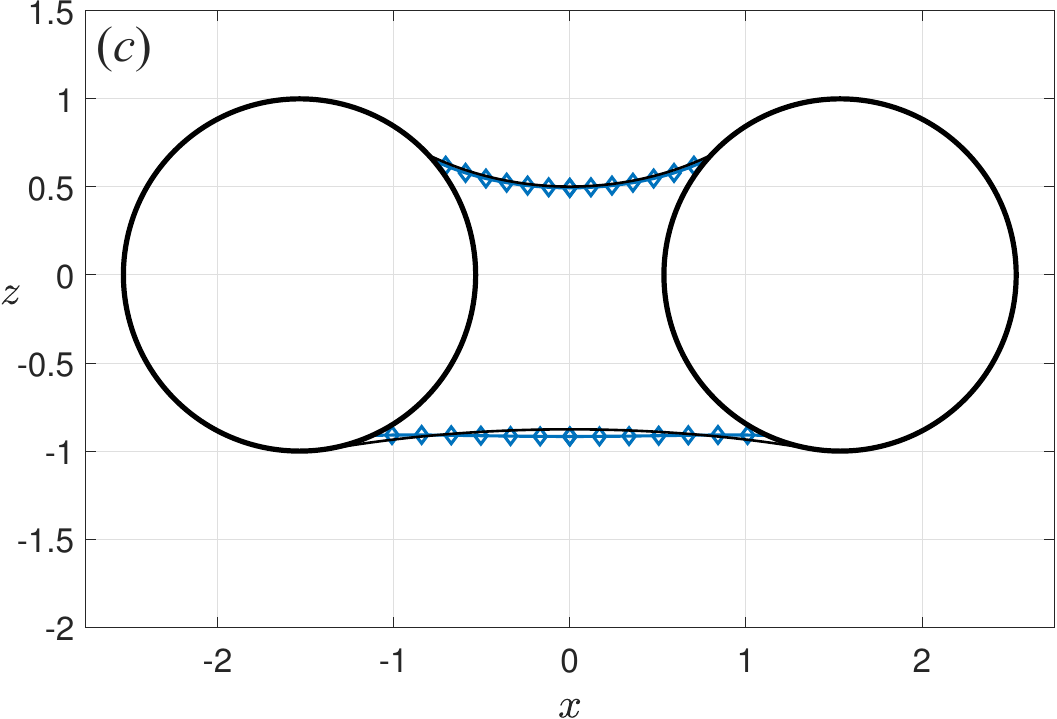}
           \includegraphics[width=0.45\textwidth]{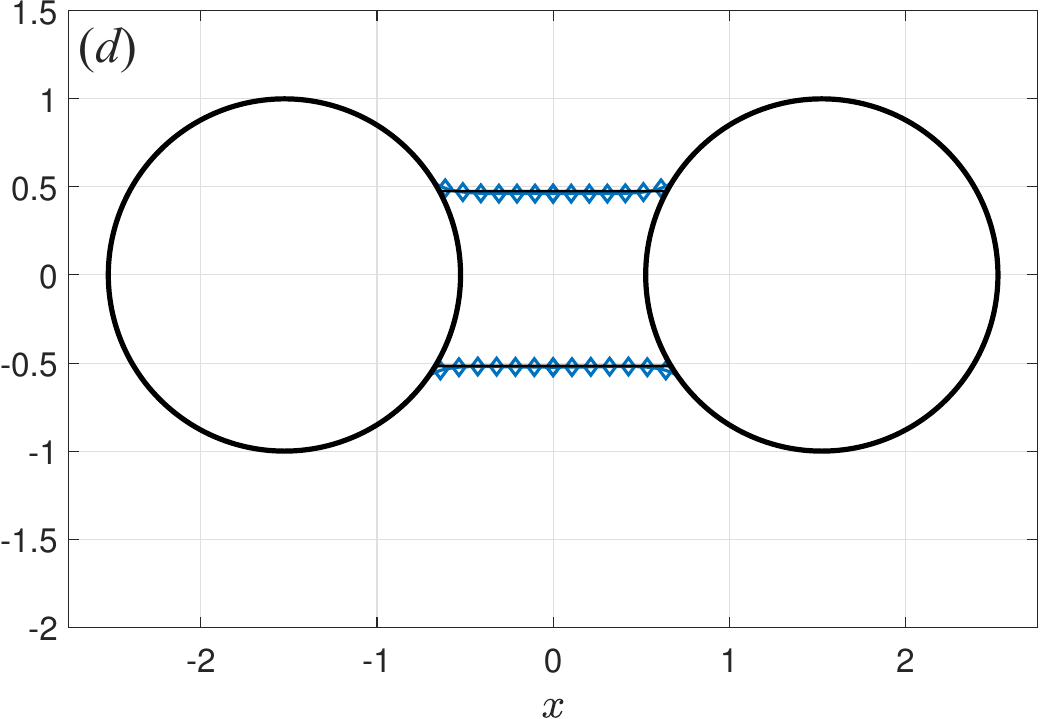}
           \caption{Experimental interface locations 
           for a $500\,\mathrm{cSt}$ silicone-oil bridge between copper rods
           compared with the eYL model solutions for different electric potential differences: (\textit{a}) non-electrified, (\textit{b}) $600\, \mathrm{V}$, (\textit{c}) $1000\, \mathrm{V}$, and (\textit{d}) $4000\, \mathrm{V}$. The corresponding geometrical and dimensionless parameters are shown in table~\ref{tab:Para3}. The physical parameters for the $500\,\mathrm{cSt}$ silicone oil are given in table~\ref{tab:Para4}.}
            \label{E_BEM}
    \end{figure}

\begin{table}
        \caption{Physical properties of the $500\,\mathrm{cSt}$ silicone oil used for the experimental results presented in figures~\ref{E_BEM}, \ref{EData6} and for the DNS results presented in figure \ref{DNS_TimeSeries}.}
	\label{tab:Para4}
	\centering
	\begin{tabular}{l l l l l}
		\toprule
        \textbf{Parameter} & \textbf{Symbol}  & \textbf{Value} \\ 
		\midrule
	 density              & $\rho$      & $970\,\, \mathrm{kg/m^3}$          \\
      surface tension      & $\gamma_{lg}$    & $0.0212\,\, \mathrm{N/m}$          \\
      kinematics viscosity & $\nu$       & $500\,\, \mathrm{cSt}$          \\
      permittivity ratio &  $\tilde{\epsilon}_l$          & $3$  \\
		\bottomrule\\
	\end{tabular}
\noindent\rule{\textwidth}{0.5pt}\\[0.2cm]
\noindent\rule{\textwidth}{0.5pt}\\[0.3cm]
     \caption{Geometrical and dimensionless parameters for the results presented in figures~ \ref{E_BEM}, \ref{EData6} and \ref{DNS_TimeSeries}.}
	\label{tab:Para3}
	\centering
	\begin{tabular}{l l l l r}
		\toprule
   \textbf{Parameter [Unit]} & \textbf{Non-Electrified}  & $\mathbf{600\,\mathrm{V}}$ & $\mathbf{1\,000\,\mathrm{V}}$ & $\mathbf{4 \,000\,\mathrm{V}}$ \\ 
		\midrule
	 $r\, [\mathrm{mm}]$         & $1$      & $1$      & $1$      & $1$      \\
      $d\, [\mathrm{mm}]$         & $0.53$ & $0.53$ & $0.53$ & $0.53$ \\
      $A\, [1]$       & $4.07$ & $2.79$ & $1.90$ & $1.10$ \\
       $\theta_0 \, [\mathrm{Rad}]$  & $0.40$  & $0.40$  & $0.40$  & $0.40$  \\
      $B\, [1]$         & $0.45$   & $0.45$   & $0.45$   & $0.45$\\
      $B_e\, [1]$       & $0$      & $0.48$   & $1.33$  & $21.25$\\
		\bottomrule\\
	\end{tabular}
\end{table}	
    
 \begin{figure}
        \centering
         \includegraphics[width=0.75\textwidth]{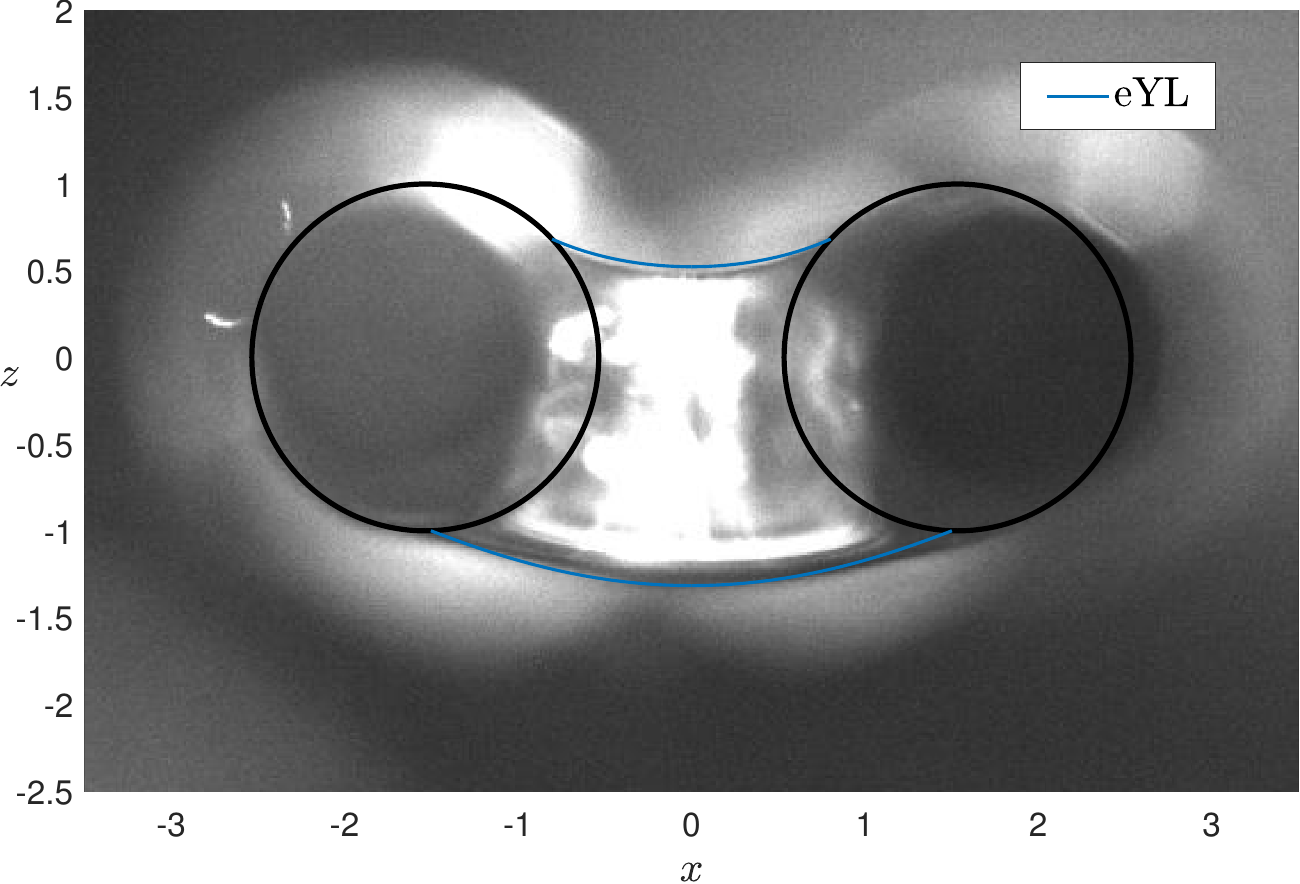}
         \caption{The solution of the the eYL model overlaid onto a photograph of a $500\,\mathrm{cSt}$ silicone-oil bridge between copper rods with a $600\,\mathrm{V}$ applied electric potential difference. The various relevant oil and rod parameters are given in tables~\ref{tab:Para3} and \ref{tab:Para4}.}
         \label{EData6}
 \end{figure}

  \begin{figure}
         \centering
         \includegraphics[width=0.75\textwidth]{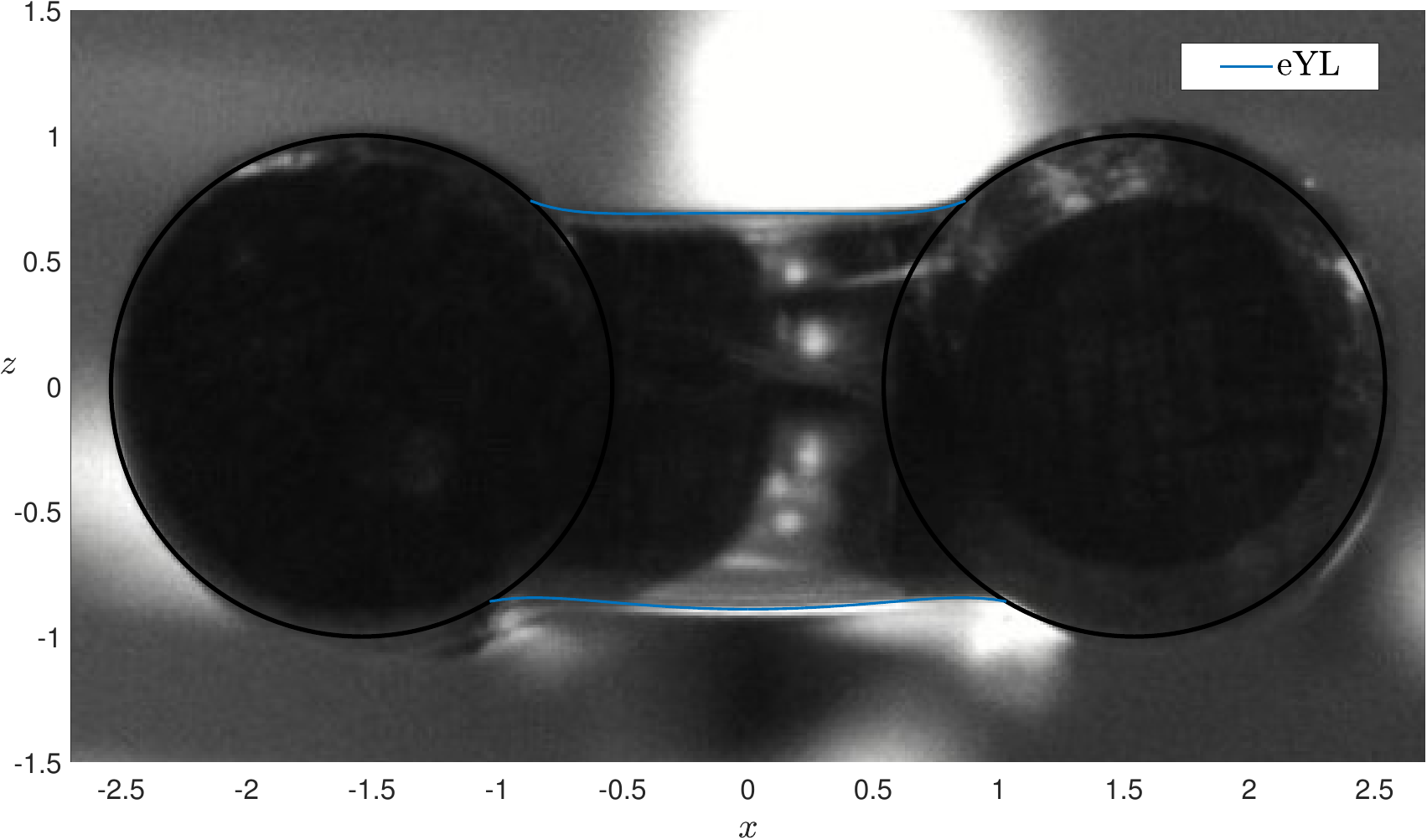}
         \caption{The solution of the the eYL model overlaid onto a photograph of a $5\,\mathrm{cSt}$ silicone-oil bridge between platinised titanium rods with a $3000\,\mathrm{V}$ applied electric potential difference. }
         \label{BEM_1V_2} 
 \end{figure}

We now compare experimental observations with steady-state solutions of the eYL model for electrified liquid bridges. 
Specifically, we compare the 2D cross sections of 3D liquid bridges at their midpoints (obtained from the experimental photos) with the corresponding eYL solutions. 
Note that in the experiments calibration measurements are first performed:
The rod diameter is measured in order to accurately extract 2D rod and liquid bridge cross sections from the experimental photos.
These measured lengths and curve shapes can then be used to determine the cross sectional area of the liquid bridge and the other relevant dimensionless parameters required for the model.

Figure~\ref{E_BEM} shows a comparison between experimentally extracted interface locations and the corresponding eYL solutions for the 2D cross-sections of $500\,\mathrm{cSt}$ silicone-oil bridges between copper rods. 
The relevant properties of the oil are given in table~\ref{tab:Para4}, and the corresponding dimensionless and geometrical parameters are listed in table~\ref{tab:Para3}.
Figure~\ref{E_BEM}(\textit{a}) corresponds to the non-electrified case, while figures~\ref{E_BEM}(\textit{b})--(\textit{d}) correspond to cases where the potential differences between the rods are $600\,\mathrm{V}$, $1000\,\mathrm{V}$ and $4000\,\mathrm{V}$, respectively.
When the electric field is applied, the liquid bridge moves upward, causing both the upper and lower interfaces to flatten.
At the same time, the cross-sectional area decreases 
due to the stretching of the liquid bridge along the rods. This is a 3D effect, the detailed analysis of which is left for future investigation.
Nonetheless, we observe that the eYL model captures the influence of the electric field very well, showing 
good agreement with the experimental results.
Figure~\ref{EData6} further illustrates this good agreement, where we overlay the eYL model prediction onto a photograph of a liquid bridge for the case of a $600\,\mathrm{V}$ potential difference. 

In figure \ref{BEM_1V_2}, we show the eYL model prediction overlaid on a photograph of a somewhat different situation: that of a silicone oil with much lower viscosity ($5\, \mathrm{cSt}$) between a pair of platinised titanium electrodes, with a potential difference of $3000\,\mathrm{V}$ 
applied between them.
The geometrical and dimensionless parameters for this case are:
$r=1\,\mathrm{mm}$, $d_0 = 0.5411\,\mathrm{mm}$, $\theta_0=20 ^{\circ}$, $A=1.0311$, $B=0.45$ and $B_e=11.95$. We again observe that the eYL model is in very good agreement with the experimental result.

\section{Dynamics of liquid bridges in the absence of an electric field}
\label{sec:dynamics}

In this section, we derive a reduced-order model for the dynamics of liquid bridges in the absence of an electric field and compare its predictions with DNS. {This provides a natural first step towards understanding the transient evolution of liquid bridges between horizontal cylinders. The extension to the fully coupled electrohydrodynamic problem 
introduces additional complexity in both modelling and DNS, and is therefore left for future work.}

\subsection{Reduced-order model based on the Onsager variational principle}
\label{sect:Onsager_Model}

The dynamics of the system in the overdamped regime (i.e.\ assuming negligible inertia) can be calculated by taking the minimum of the Rayleighian according to the Onsager variational principle, see e.g.\ \cite{DoiSoft}, \cite{LopesThiele}, \cite{DoiMan}. The Rayleighian is
\begin{equation}
	 \mathcal{R}= \dot{F}+\Phi,
\end{equation}
where $\dot{F}\equiv d F/dt$ is the time derivative of the free energy and $\Phi$ is the dissipation function.
Note that more generally the dissipation function and the Rayleighian are both functionals, but in the present case we simplify so that they are functions of just a few state variables, which we introduce below. 

We consider the non-electrified case, for which the free energy, $F$, is given by 
\begin{equation}
 \label{lambdaA}
     F=\mathcal{F}+G-\lambda ({\cal A}-A_0).
\end{equation}
Here, $\mathcal{F}$ is the interfacial free energy, $G$ is the gravitational potential energy and ${\cal A}$ represents the area of the cross section of the bridge (which depends on the state of the system), with $\lambda$ denoting a Lagrange multiplier, which is identified with the pressure $p_0$ in \eqref{Y_Laplace11} and \eqref{Y_Laplace22}. 
The last term in (\ref{lambdaA}) is introduced to ensure conservation of volume, so that ${\cal A}=A_0$, where $A_0$ is the specified cross-sectional area of the bridge.
Note that we do not incorporate the effects of any electric fields into the Onsager model.
It would be at this stage, via the free energy, that such contributions could be included, but we defer doing this to future work.

\begin{figure}
    \centering
    \includegraphics[width=0.7\textwidth]{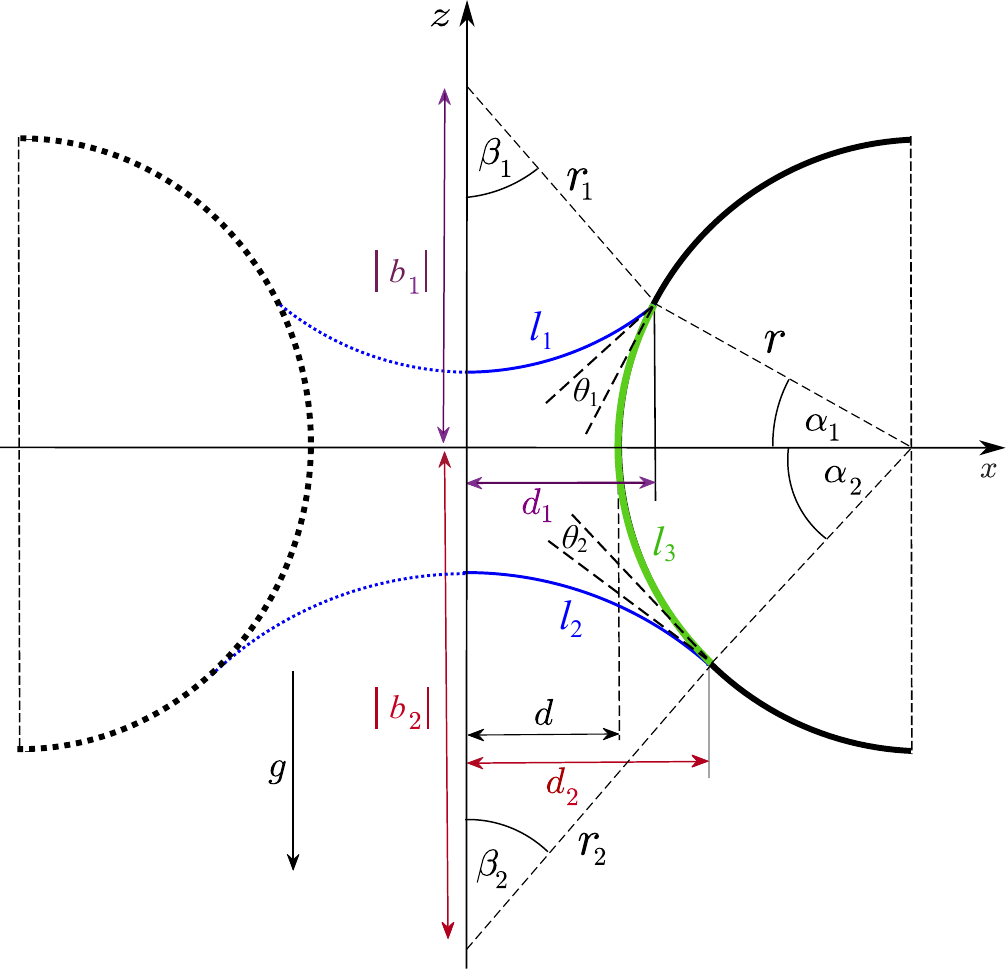}
    \caption{2D sketch of the liquid bridge under the assumption of the circular interfaces indicating various parameters for this geometry.}
	\label{a_figure2zoom}
\end{figure}

The interfacial free energy is given by
\begin{equation}
\label{surfaceE}
    \mathcal{F}=  \, 2 \,l_1 \gamma_{lg}+2 \,l_2 \gamma_{lg} 
                 + 2\,l_3\, (\gamma_{sl}-\gamma_{sg}),               
\end{equation}
which is composed of several terms each consisting of an interfacial tension multiplied by the corresponding length of that interface.
We recall that $l_1$ and $l_2$ are the half-lengths of the upper and lower liquid--air interfaces, respectively (see figure~\ref{a_figure2zoom}), and $l_3$ is the full length of the wetted part of the right cylinder. The factor $2$ in each of the terms in \eqref{surfaceE} is due to the symmetry assumption, introduced earlier. Also, $\gamma_{lg}$, $\gamma_{sl}$ and $\gamma_{sg}$ are the liquid--air, solid--liquid and solid--air surface tensions, respectively.

The gravitational potential energy is
\begin{equation}
  G={\cal A} \, \rho \, g\, \bar{z},
\end{equation}
 where $\rho$ is the uniform density of the liquid and  $g$ is the gravitational acceleration. Furthermore, $\bar{z}$ is the $z$ coordinate of the centre of mass of the liquid bridge, which can be calculated as 
$	\bar{z}=\left.\int_{\mathcal{W}_l} z \, \rho \, dA\right/ {\cal A}\, \rho$, where $\mathcal{W}_l$ denotes the cross section of the liquid bridge. Therefore, the potential energy is 
\begin{equation}
\label{potential_energy}
  G=
  {\cal A} \, \rho \, g \, \frac{\int_{\mathcal{W}_l} z \, \rho \, dA}{{\cal A}\,\rho} =
  \rho \, g\int_{\mathcal{W}_l}  z \,  dA.
\end{equation}

Note that minimisation of the free energy, $F$, given by \eqref{lambdaA} with $\mathcal{F}$ and $G$ defined by \eqref{surfaceE} and \eqref{potential_energy}, respectively, leads to the equations for the interfacial shapes \eqref{Y_Laplace11} and \eqref{Y_Laplace22} (with $\mathcal{E}_1=0$ and $\mathcal{E}_2=0$, as we consider the non-electrified case here). In addition, the transversality condition at each of the three-phase contact lines simply implies that the equilibrium contact angle that the liquid makes with the cylinder surface, $\theta_0$, satisfies Young's equation 
\begin{equation}\label{eq:Youngs}
    \cos \theta_0 = \frac{(\gamma_{sg}-\gamma_{sl})}{\gamma_{lg}}.
\end{equation}

To obtain a reduced-order model for the dynamics, we approximate the shapes of the interfaces by arcs of circles. Note that in the absence of gravity, the interfaces are indeed arcs of circles, which we can show analytically.
Hence, we expect that this assumption remains appropriate for small liquid bridges.   
Under this assumption, the state of the system is fully determined by four angles: 
the angles to the horizontal $\alpha_1$ and $\alpha_2$ that determine the points of contact between the right cylinder and the upper and lower interfaces, respectively, and  the contact angles $\theta_1$ and $\theta_2$ between the cylinder and the top and bottom interfaces, respectively -- see figure~\ref{a_figure2zoom}. So under this assumption, the state variable is  
\refstepcounter{equation}
\begin{equation}
    \bm{\xi} = (   \alpha_1, \alpha_2, \theta_1, \theta_2 )^T. 
           \quad 
\end{equation}
Computing the arc lengths $l_i$, $i=1,2,3$, and using Young's equation (\ref{eq:Youngs}), leads to the following expression for the interfacial free energy: 
\begin{equation}
\begin{split}
 	\mathcal{F}(\bm{\xi})=
 	& \,2 \left(\pi/2-\alpha_1-\theta_1\right) r_{1} \gamma_{lg} 
        + 2 \left(\pi/2-\alpha_2-\theta_2 \right) r_{2} \gamma_{lg} \\
        &-2 \gamma_{lg} (\cos{\theta_0}) r \alpha_1 
 	-2 \gamma_{lg} (\cos{\theta_0}) r \alpha_2.
\end{split} 	
\end{equation} 
Here, $r_{1}$ and $r_{2}$ are the radii of the upper and lower circular interfaces, given by
\refstepcounter{equation}
$$
 	r_{1}=\frac{r+d-r\cos\alpha_1}{\cos \left( \theta_1+\alpha_1\right)}, \quad \quad \quad
 	r_{2}=\frac{r+d-r\cos\alpha_2}{\cos \left( \theta_2+\alpha_2\right)} .
        \eqno{(\theequation{\mathit{a},\mathit{b}})}
$$
The gravitational potential energy under these assumptions takes the form
\begin{equation}
\begin{split}
 	G(\bm{\xi})=
 	&\rho g \biggl[ \left(b_1^2+r_{1}^2+d^2+2rd \right)d_1-\left(d+r\right)d_1^2 
    \\ &-b_1\left(d_1\sqrt{r_{1}^2-d_1^2}-r_{1}^2\arctan{\frac{d_1\sqrt{r_{1}^2-d_1^2}}{d_1^2-r_{1}^2}}\right)	
\\&-\left(b_{2}^2+r_{2}^2+d^2+2rd\right)d_2+\left(d+r\right)d_2^2
 \\& -b_2\left(d_2\sqrt{r_{2}^2-d_2^2}-r_{2}^2\arctan{\frac{d_2\sqrt{r_{2}^2-d_2^2}} {d_2^2-r_{2}^2}}\right) \biggr],
\end{split} 	
\end{equation} 
where $d_1$ and $d_2$ are the $x$ coordinates of the three-phase contact points at the right cylinder (on the upper and lower interfaces, respectively) given by
\begin{align*} 
   d_1=d+r-r \cos{\alpha_1}, \quad d_2=d+r-r \cos{\alpha_2}.
   \refstepcounter{equation}\tag{\theequation \textit{a,b}} \label{a1a2} 
\end{align*}
Also, $b_1$ and $b_2$ are the $z$ coordinates of the centres of the circles determining the upper and lower liquid--air interfaces, respectively (see figure \ref{a_figure2zoom}), given by
\begin{align*} 
 b_1=r_{1} \cos{\beta_1}+r \sin{\alpha_1}, \quad b_2=-r_{2} \cos{\beta_2}-r \sin{\alpha_2}.
 \refstepcounter{equation}\tag{\theequation \textit{a,b}} \label{b1b2} 
\end{align*}
With the assumption of circular interfaces, the area is fully determined by the four angles (considering $r$ and $d$ as parameters), and it can be shown that
\begin{align}
\begin{split}	
        {\cal A}(\bm{\xi}) =
        &\,2 r_{1} r \sin \beta_1 \sin\alpha_1-r^2 \left( \alpha_1-\sin\alpha_1\cos\alpha_1\right) \\
        &-r_{1}^2\left( \beta_1-\sin\beta_1\cos\beta_1\right)+2r_{2}r\sin\beta_2\sin\alpha_2 \\
        &-r^2\left(\alpha_2-\sin\alpha_2\cos\alpha_2 \right)-r_{2}^2 \left(  \beta_2-\sin\beta_2\cos\beta_2\right),	
\end{split}
\end{align}
where $\beta_1=\pi/2-\theta_1-\alpha_1$ and $\beta_2=\pi/2-\theta_2-\alpha_2$.

For the energy dissipation function, as usual we use a quadratic approximation \citep{DoiSoft},  
\begin{equation}
    \Phi(\dot{\bm{\xi}})=\frac{1}{2}\, \dot{\bm{\xi}} \cdot \bm{D} \cdot \dot{\bm{\xi}}, 
\end{equation}
where $\bm{D}\equiv(d_{ij})_{1\leq i,j \leq 4}$ is a symmetric positive definite friction coefficient or dissipation matrix. 

In order to compute the state evolution equation, we minimise the Rayleighian function with respect to the rates $\dot{\alpha_1}$, $\dot{\alpha_2}$, $\dot{\theta_1}$ and $\dot{\theta_2}$. 
Using that the rate of change of the free energy~is
\begin{equation}
    \dot{F}(\bm{\xi})=\frac{\partial F}{\partial \alpha_1} \dot{\alpha_1}+
                       \frac{\partial F}{\partial \alpha_2} \dot{\alpha_2}+
                       \frac{\partial F}{\partial \theta_1} \dot{\theta_1}+
                       \frac{\partial F}{\partial \theta_2} \dot{\theta_2},
\end{equation}
the dynamics is given by he following system of coupled equations:
\begin{subeqnarray}
	\frac{\partial \mathcal{R}}{\partial \dot{\alpha_1}} & = & \frac{\partial F}{\partial \alpha_1}+ d_{11} \dot{\alpha_1}+ d_{12} \dot{\alpha_2} + d_{13} \dot{\theta_1}+ d_{14} \dot{\theta_2} = 0, \\
	\frac{\partial \mathcal{R}}{\partial \dot{\alpha_2}} & = & \frac{\partial F}{\partial \alpha_2}+ d_{12} \dot{\alpha_1}+ d_{22} \dot{\alpha_2} + d_{23} \dot{\theta_1}+ d_{24} \dot{\theta_2} = 0, \\
	\frac{\partial \mathcal{R}}{\partial \dot{\theta_1}} & = & \frac{\partial F}{\partial \theta_1}+ d_{13} \dot{\alpha_1}+ d_{23} \dot{\alpha_2} + d_{33} \dot{\theta_1}+ d_{34} \dot{\theta_2} = 0, \\
	\frac{\partial \mathcal{R}}{\partial \dot{\theta_2}} & = & \frac{\partial F}{\partial \theta_2}+ d_{14} \dot{\alpha_1}+ d_{24} \dot{\alpha_2} + d_{34} \dot{\theta_1}+ d_{44} \dot{\theta_2} = 0.
\end{subeqnarray}  
Here, we have exploited the fact that the dissipation matrix $\bm{D}$ is symmetric. 

Using that $(\nabla A)^T  \dot{\bm{\xi}}=0$ (which follows from volume conservation), the dynamics equations can be written as
\begin{equation}
    \dot{\bm{\xi}}= -\bm{D}^{-1}(\nabla (\mathcal{F}+G)-\lambda \nabla {\cal A}),
\end{equation}
where 
\begin{equation}
    \nabla=\left(\frac{\partial }{\partial \alpha_1},  \frac{\partial }{\partial \alpha_2},  \frac{\partial }{\partial \theta_1},  \frac{\partial }{\partial \theta_2} \right)^T,
\end{equation}
and the Lagrange multiplier $\lambda$ is given by 
\begin{equation}  
\lambda = \frac{(\nabla {\cal A})^T  \bm{D}^{-1} \nabla (\mathcal{F}+G)} { (\nabla {\cal A})^T \bm{D}^{-1} \nabla {\cal A}}.
\end{equation}

\begin{figure}
    \centering
    \includegraphics[width=0.6 \textwidth]{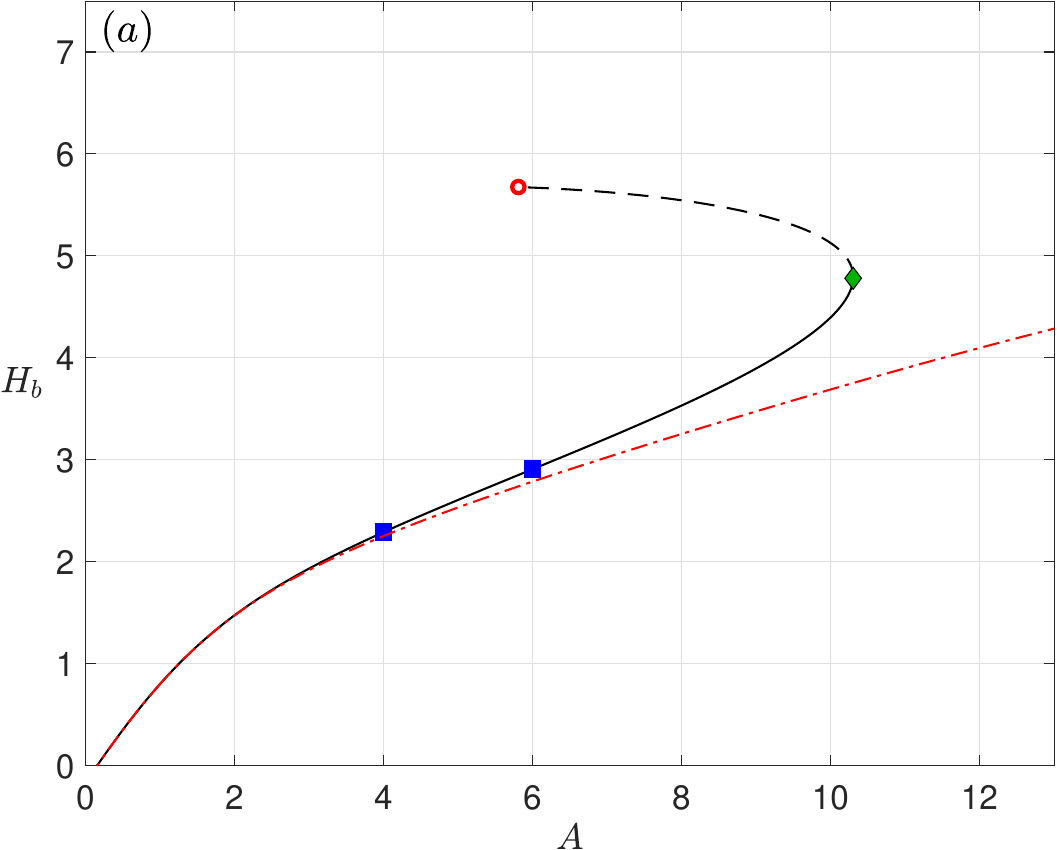}\\[0.3cm]
    \includegraphics[height=5.3cm]{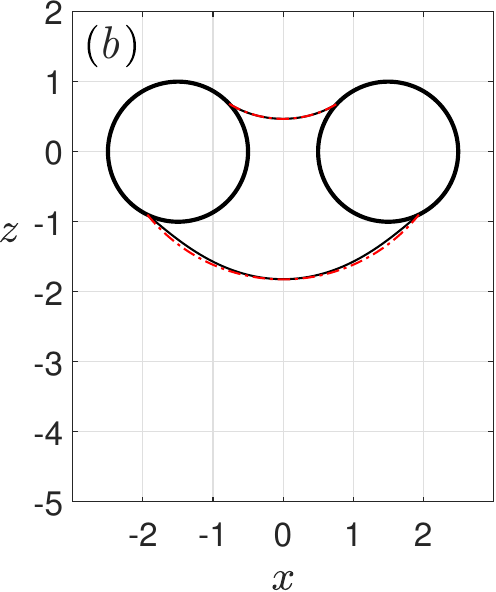}
    \includegraphics[height=5.3cm]{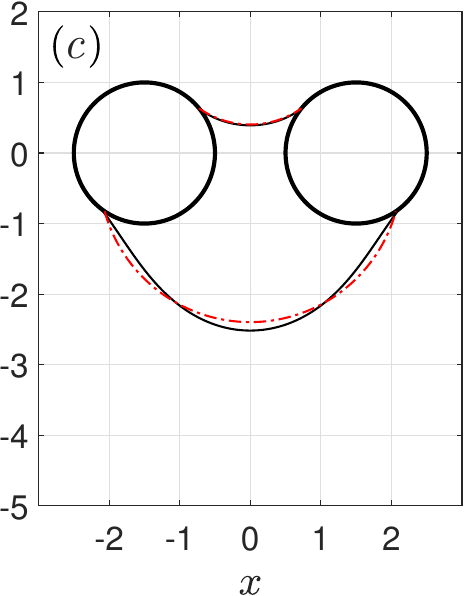}
    \includegraphics[height=5.3cm]{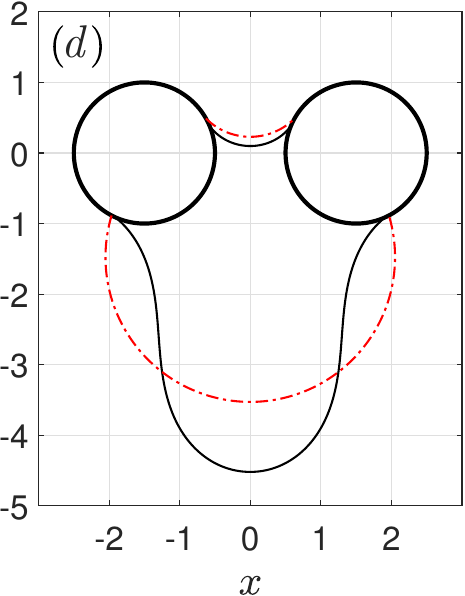}
    \caption{(a) The bridge thickness in the middle $H_b$, varying with the area of the bridge cross-section $A$, for the same parameter values as in figure~\ref{V_0_1}. The black solid and dashed lines indicate stable and unstable solutions, respectively, of the full governing equations, while the red dash-dotted line indicates steady-state solutions of the Onsager model. The blue filled squares correspond to $A=4$ and $A=6$, and the respective solutions are shown in panels (\textit{b}) and (\textit{c}). The green filled diamond corresponds to the maximal trapping capacity $A=10.3$ and the respective solution is shown in panel (\textit{d}). As in figure~\ref{V_0_1}, the red empty circle corresponds to the unstable solution where the profile for the lower liquid--air interface pinches.
    }
	\label{compare_02}
\end{figure}

\subsection{Steady-state validation of the Onsager model}
\label{subsec:8.1}

Before discussing the dynamics predicted by the reduced-order Onsager model, we first compare steady-state solutions of this model with those of the full governing equations for the non-electrified case (i.e.\ with solutions of the boundary-value problem described in \S~\ref{Sec:Steady_nonelectrified}). We consider the same values of the dimensionless parameters as in table~\ref{tab:Para1}, but with $B_e=0$ (i.e. without electric field). In figure~\ref{compare_02}(\textit{a}), we plot the dimensionless liquid bridge thickness $H_b$ versus the dimensionless cross-sectional area $A$. As before (see figure~\ref{V_0_1}(a)), the black solid and dashed lines correspond to stable and unstable solutions, respectively, of the full governing equations. The green filled diamond indicates the turning point at $A=10.3$, which is the maximal cross-sectional bridge area. The red empty circle corresponds to the point where the solution profile for the lower liquid--air interface pinches. The red dash-dotted line  corresponds to steady-state solutions of the reduced-order Onsager model.
Unlike for the full model, this line continues without a turning point beyond the maximal trapping capacity $A=10.3$. However, we see very good agreement between the results from the Onsager model and from the full governing equations for smaller values of $A$.
In fact, the red dash-dotted line is nearly on top of the black solid line for $A\lesssim 4$.
The steady-state solutions for $A=4$ are shown in figure~\ref{compare_02}(\textit{b}). 
For larger values of $A$, discrepancies between the solutions increase. An example is shown in figure~\ref{compare_02}(\textit{c}) for $A=6$. There are visible differences between the bottom liquid--air interfaces for the full and reduced-order models, while the agreement remains very good between the top liquid--air interfaces for the two models. The differences between the solutions for the two models further increase as $A$ increases. Figure~\ref{compare_02}(\textit{d}) compares the solutions at the maximal trapping capacity for the full governing equations, i.e. at $A=10.3$. The lower interfaces differ significantly and we now also observe noticeable disagreement between the top interfaces too. 

We conclude that the model provides good predictions for moderate values of $A$ (${A\lesssim 4}$), but becomes less accurate for larger $A$, where the assumption of circular interface shapes is less well satisfied due to stronger gravity-induced deformations of the interfaces, particularly of the lower one.

\begin{figure}
    \centering
    \hspace{0.4cm}\includegraphics[height=5.0cm]{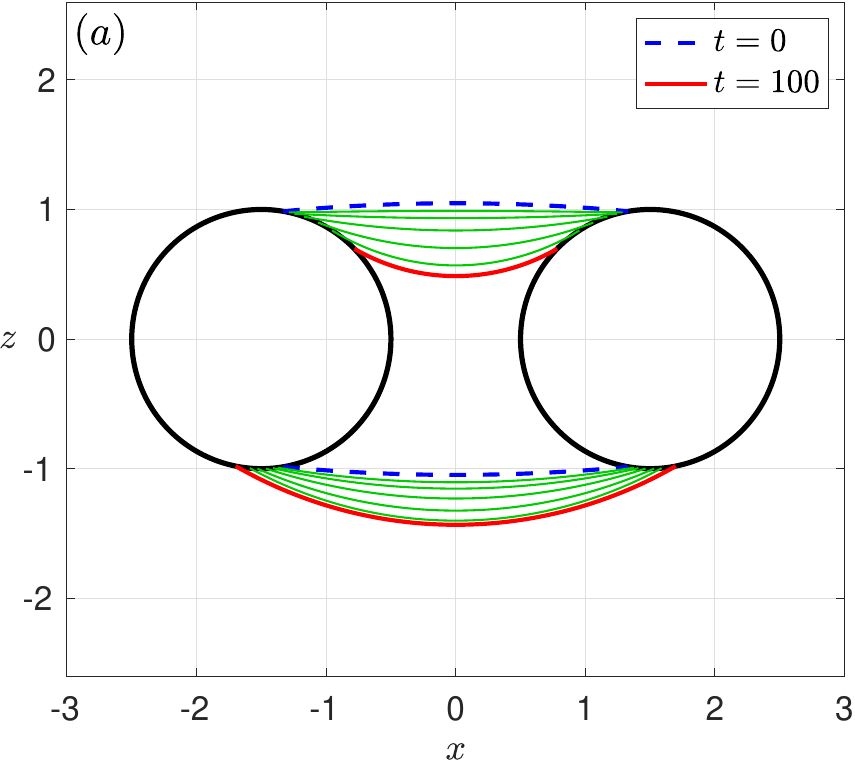}\hspace{0.4cm}
    \includegraphics[height=5.0cm]{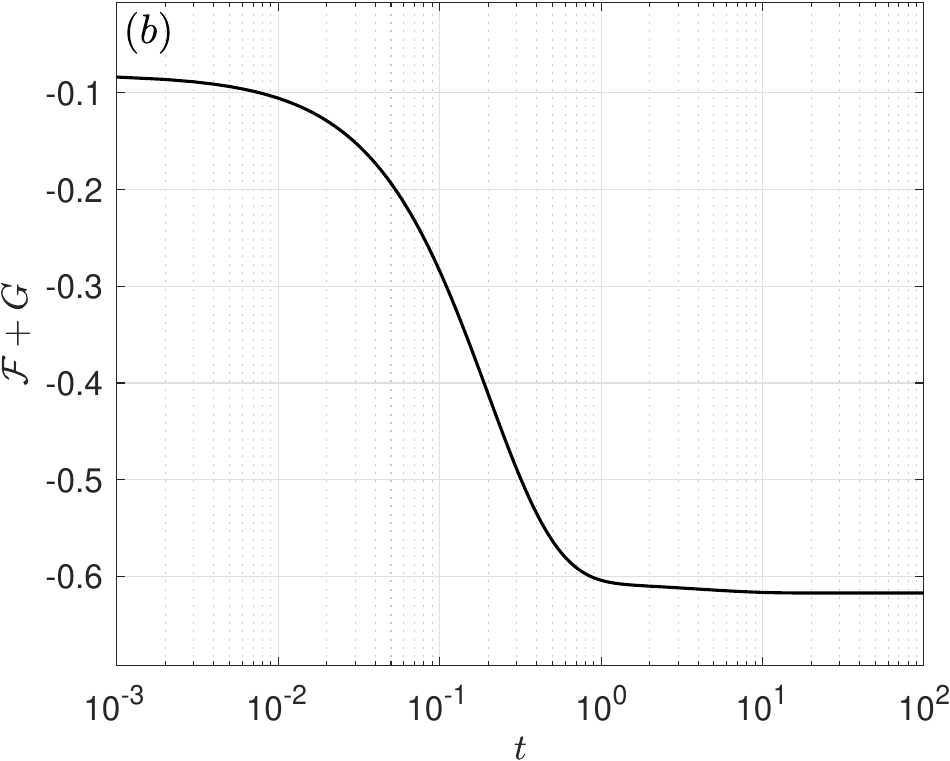}\\[0.3cm]
    \includegraphics[height=5.0cm]{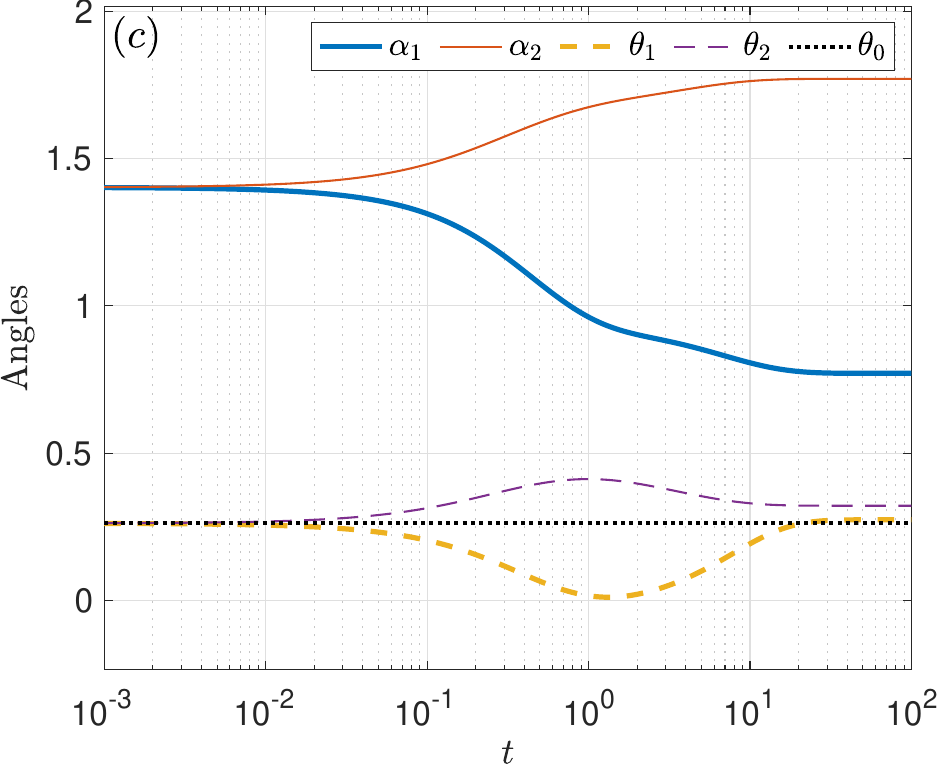}\hspace{0.2cm}
    \includegraphics[height=5.0cm]{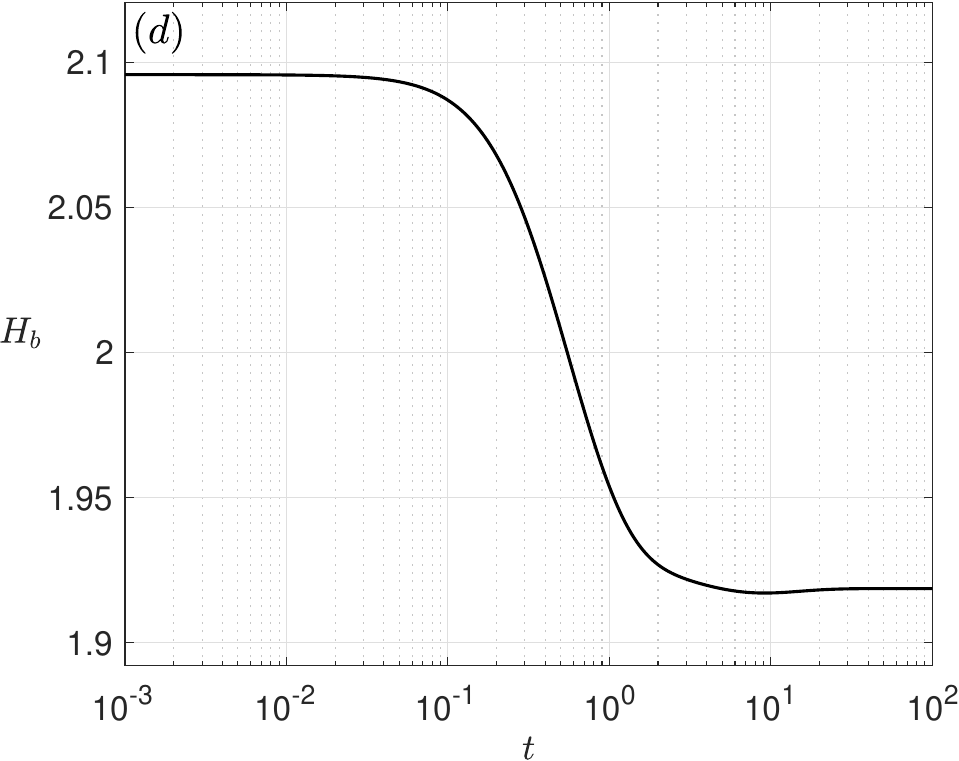}    
    \caption{The dynamics of the liquid--air interfaces obtained using the reduced-order Onsager model. The values of the dimensionless parameters are the same as in figure~\ref{V_0_1} (but with $B_e=0$), and the dimensionless cross-sectional area of the bridge is $A=3$. Panel (\textit{a}) shows locations of the interfaces at different times.  The initial interfaces are shown with the blue dashed lines. The near-equilibrium state at $t=100$ is show with the red thick solid lines. The green thin solid lines indicate intermediate liquid--air interface positions at $t=0.05$, $0.1$, $0.2$, $0.4$ and $0.8$. Panel (\textit{b}) shows time evolution of the energy $\mathcal{F}+G$. Panel (\textit{c}) shows time evolution of the upper and the lower contact angles $\theta_1$ and $\theta_2$ and the angles $\alpha_1$ and $\alpha_2$ that determine the upper and lower three-phase contact points. The equilibrium contact angle $\theta_0$ is also indicated. The angles are given in radians. Panel (\textit{d}) shows time evolution of the bridge thickness $H_b$.}
	\label{evolution_A_3}
\end{figure}

\begin{figure}
    \centering
    \hspace{0.4cm}\includegraphics[height=5.0cm]{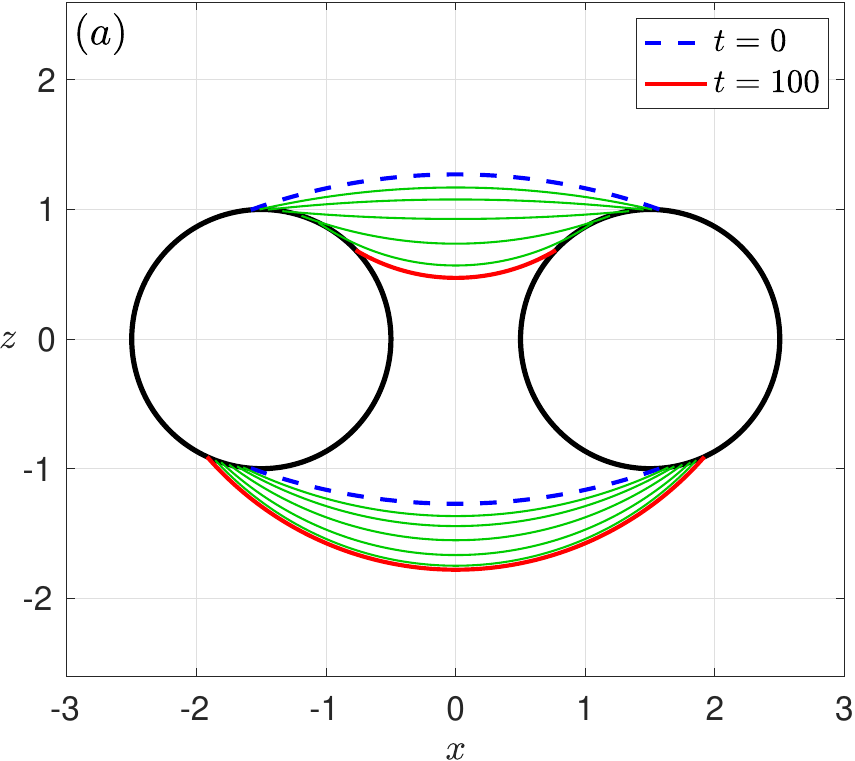}\hspace{0.4cm}
    \includegraphics[height=5.0cm]{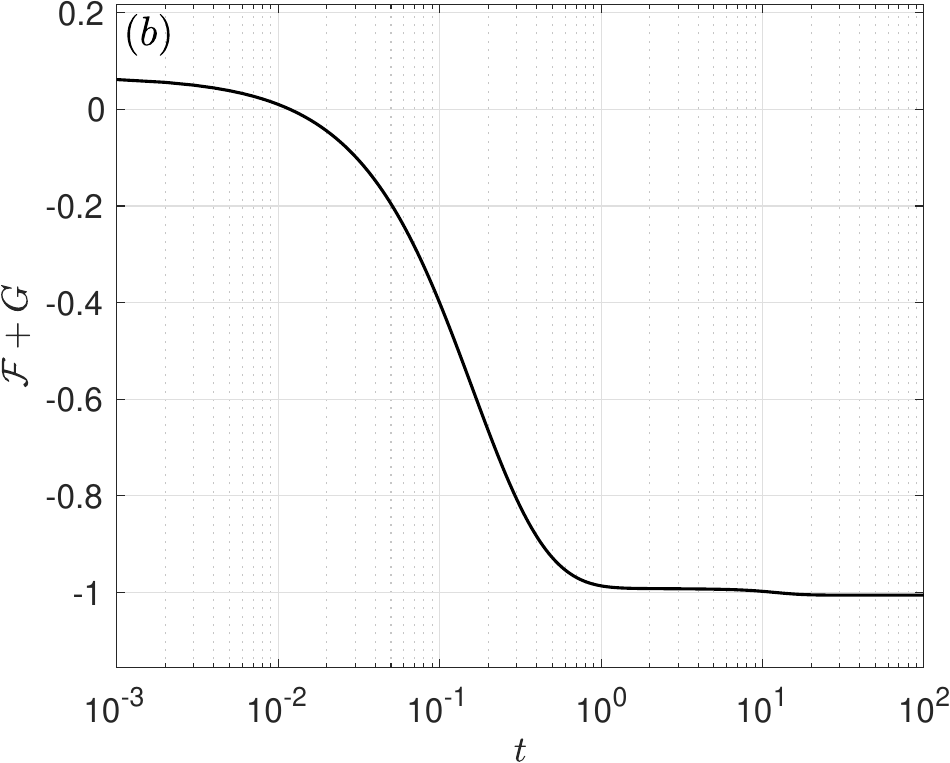}\\[0.3cm]
    \includegraphics[height=5.0cm]{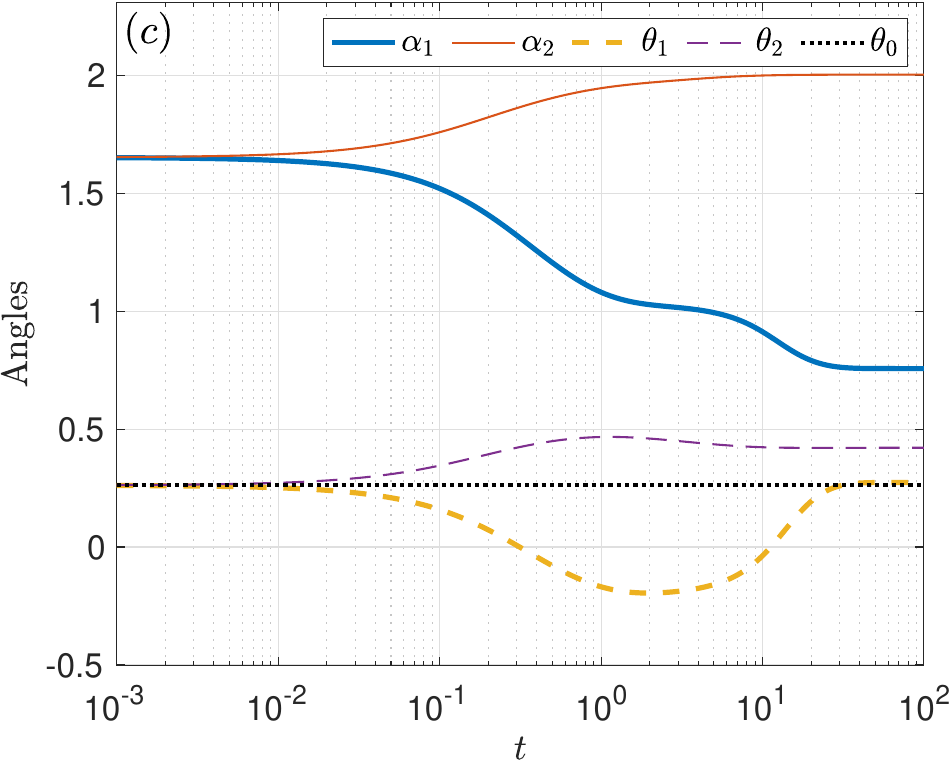}\hspace{0.2cm}
    \includegraphics[height=5.0cm]{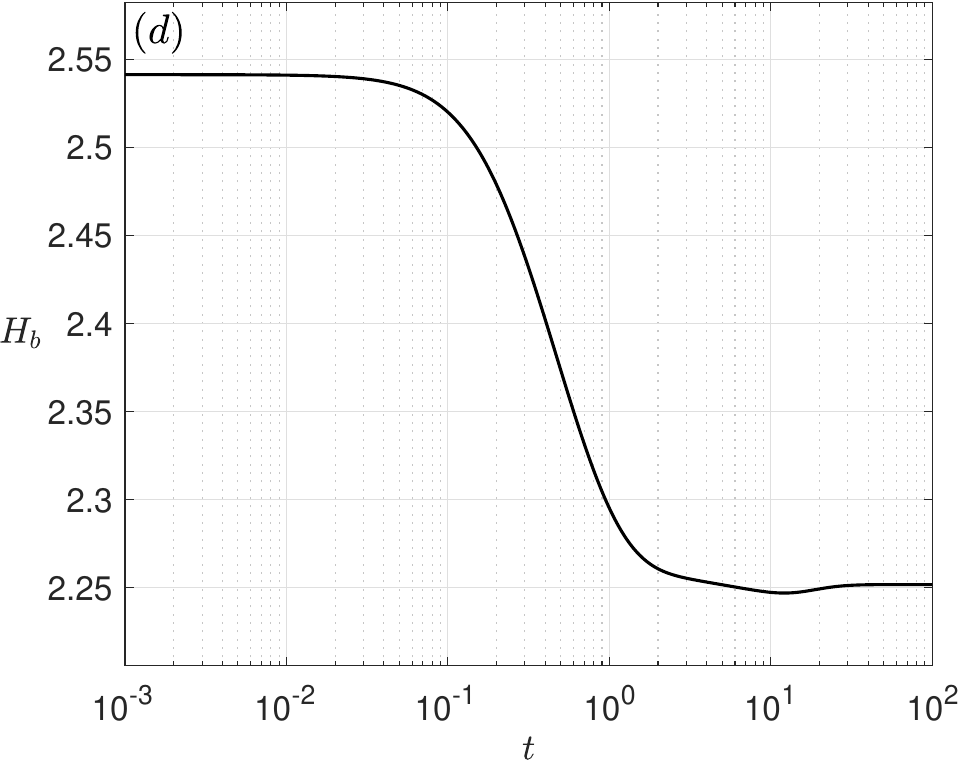}
    \caption{This is the same as figure \ref{evolution_A_3}, except here $A=4$.}
	\label{evolution_A_4}
\end{figure}

\subsection{Dynamics predicted by the Onsager model}
\label{subsec:onsager_dynamics}

We now consider the evolution of liquid bridges using the reduced-order Onsager model.
Examples are shown in figure~\ref{evolution_A_3} (for $A=3$) and in figure~\ref{evolution_A_4} (for $A=4$), where the same parameter values as in figures~\ref{V_0_1}(\textit{a},\textit{b}) 
are used. For these simulations, we set the dissipation matrix, $\boldsymbol{D}$, to be equal to the identity matrix, with the overall prefactor (a mobility coefficient) being absorbed into the timescale.
This coefficient can be approximated by matching with either the experimental results or the results of DNS. We return to this below.
The initial locations of the liquid--air interfaces are indicated by the blue dashed lines in panels (\textit{a}) of figures~\ref{evolution_A_3} and \ref{evolution_A_4} and these correspond to the steady-state liquid bridges in the absence of gravity for $A=3$ and $A=4$, respectively. The red lines correspond to the liquid--air interfaces at time $t=100$, at which the liquid bridges have practically reached the final equilibrium states. Intermediate liquid--air interface positions at  $t=0.05$, $0.1$, $0.2$, $0.4$ and $0.8$ are shown with the green lines in panels (\textit{a}).

Panels (\textit{b}) of figures~\ref{evolution_A_3} and \ref{evolution_A_4} show the evolution of the sum of the interfacial energy $\mathcal{F}$ and the gravitational potential energy $G$. They confirm that $\mathcal{F}+G$ is monotonically decreases, as expected. Panels (\textit{c}) of figures~\ref{evolution_A_3} and \ref{evolution_A_4} show the evolution of the state variables for the Onsager model, i.e.\ the evolution of the upper and the lower contact angles $\theta_1$ and $\theta_2$ and the angles $\alpha_1$ and $\alpha_2$ that determine the upper and lower three-phase contact points. The equilibrium contact angle $\theta_0$ is also shown as the horizontal line, and it has been used as the initial value for both $\theta_1$ and $\theta_2$. It can be observed that both for $A=3$ and for $A=4$ the upper contact angle, $\theta_1$, initially decreases to a minimum value 
and then monotonically increases and converges to the values $0.2737$ and $0.2738$ for $A=3$ and $A=4$, respectively, 
which are close to the equilibrium contact angle $\theta_0=\pi/12\approx 0.2618$. The lower contact angle, $\theta_2$, initially monotonically increases to a maximum value and then converges to the values $0.321$ and $0.421$ for $A=3$ and $A=4$, respectively, which differ significantly from $\theta_0$. The fact that the contact angles $\theta_1$ and $\theta_2$ for the equilibrium state of the Onsager model differ from $\theta_0$ can be explained as follows:
Although the assumption of circular shapes of the liquid--air interfaces proves to be good for moderately small  values of $A$, it is, strictly speaking, not valid in the presence of gravity.
It turns out that the minimum state for the reduced free-energy function is obtained for $\theta_1$ and $\theta_2$ different from $\theta_0$. However, this state still gives good overall agreement for the liquid--air interfaces with the steady-state solution profiles for the full governing equations for a wide range of $A$ values (for $A\lesssim 4$, as already mentioned).
Note, however, that while for $A=3$ the upper contact angle remains positive for all $t$ (nearly reaching zero at its minimum point), for $A=4$ the upper contact angle becomes negative for $t\in(0.301,11.248)$.
This is of course non-physical, and indicates some failure of the model. We conclude that although the reduced-order model gives good prediction of steady-state solutions for $A\lesssim4$, it is only valid for $A\lesssim 3$, as far as the dynamics is concerned. 
To improve the reduced-order model, it must be extended by having more degrees of freedom in the shapes of the liquid--air interfaces.
This is left as a topic for future investigation.
Finally, we note that panels (\textit{d}) of figures~\ref{evolution_A_3} and \ref{evolution_A_4} show the evolution of the bridge thickness $H_b$, for $A=3$ and $A=4$, respectively. In both cases, $H_b$ initially decreases monotonically to its minimum value before converging to the equilibrium state.

\subsection{Comparison with direct numerical simulations}
\label{subsec:DNS}

To complement our benchmark experimental results in \S~\ref{sec:2} and further validate the reduced-order Onsager model, 
which enables a highly efficient exploration of the parameter space, we perform DNS for the non-electrified case within the  open-source VOF Basilisk framework \citep{popinet2009accurate,popinet2015quadtree}. The DNS  also allow the dynamics to be examined in regimes where the Onsager model begins to break down, and provide access to details that are difficult to resolve experimentally. The details of the DNS setup are discussed in Appendix~\ref{sec:DNS}. 

A fully coupled electrohydrodynamic formulation is not included in the DNS.  While such simulations have been performed in Gerris/Basilisk and other frameworks \cite[see e.g.][]{tomar2007,lopez2011,lin2012,esmaeeli2020}, they typically involve simpler geometries. In the present configuration, the presence of curved embedded solids and moving contact lines makes a fully coupled formulation significantly more involved. We therefore restrict our attention to the non-electrified case, leaving the DNS implementation of the fully coupled multi-physics problem for future work.

     \begin{figure}
         \centering
            \includegraphics[width=0.44\textwidth]{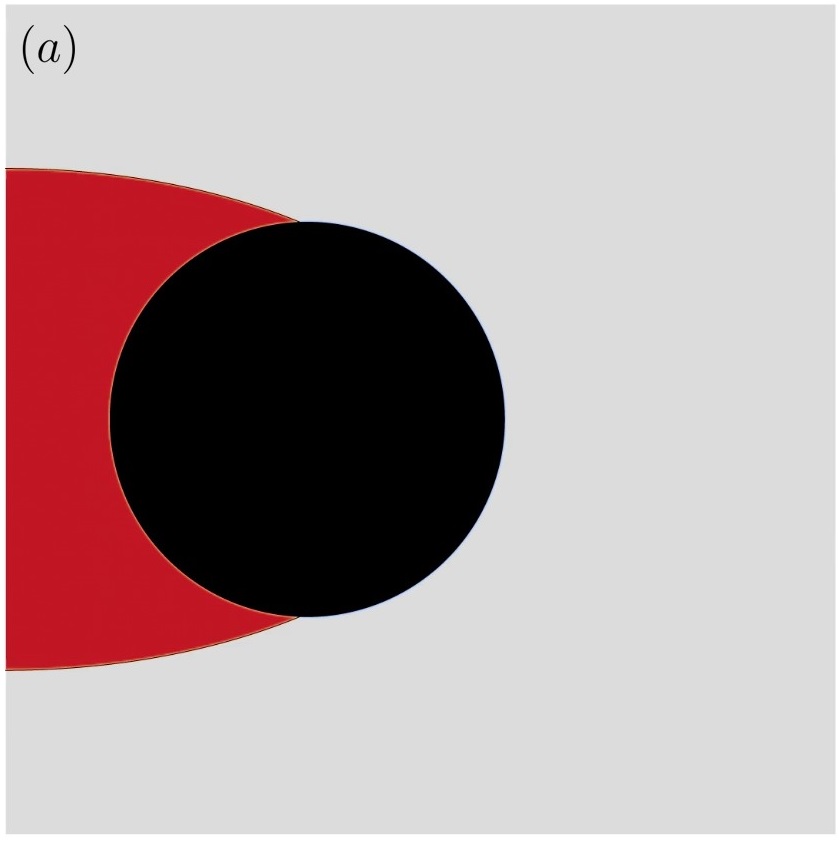}
            \includegraphics[width=0.44\textwidth]{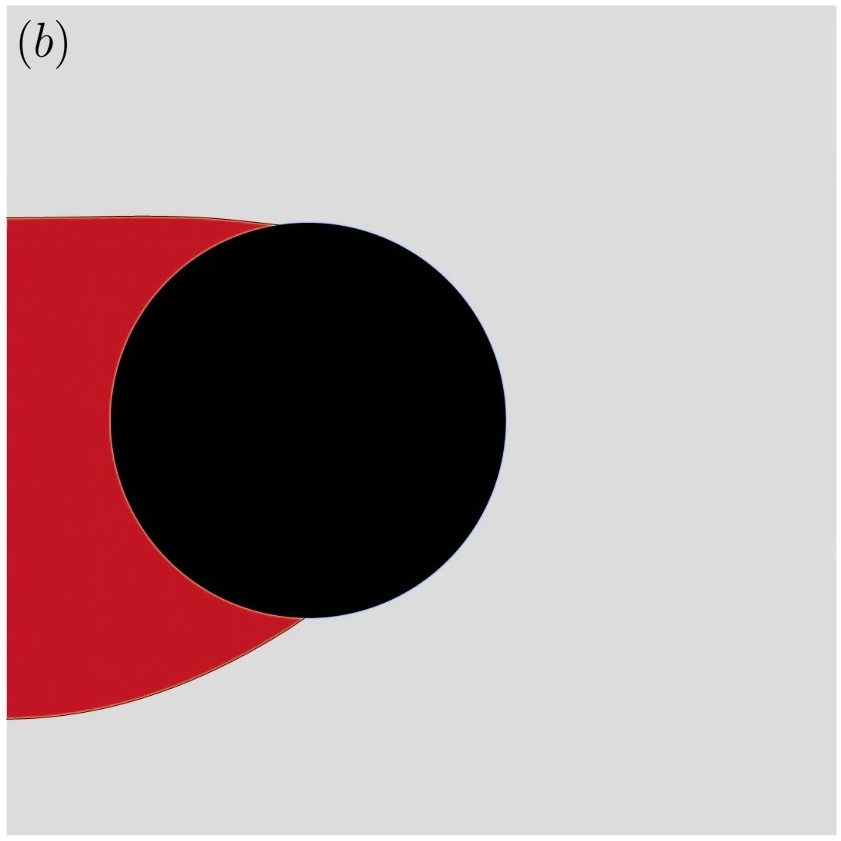}
            \includegraphics[width=0.44\textwidth]{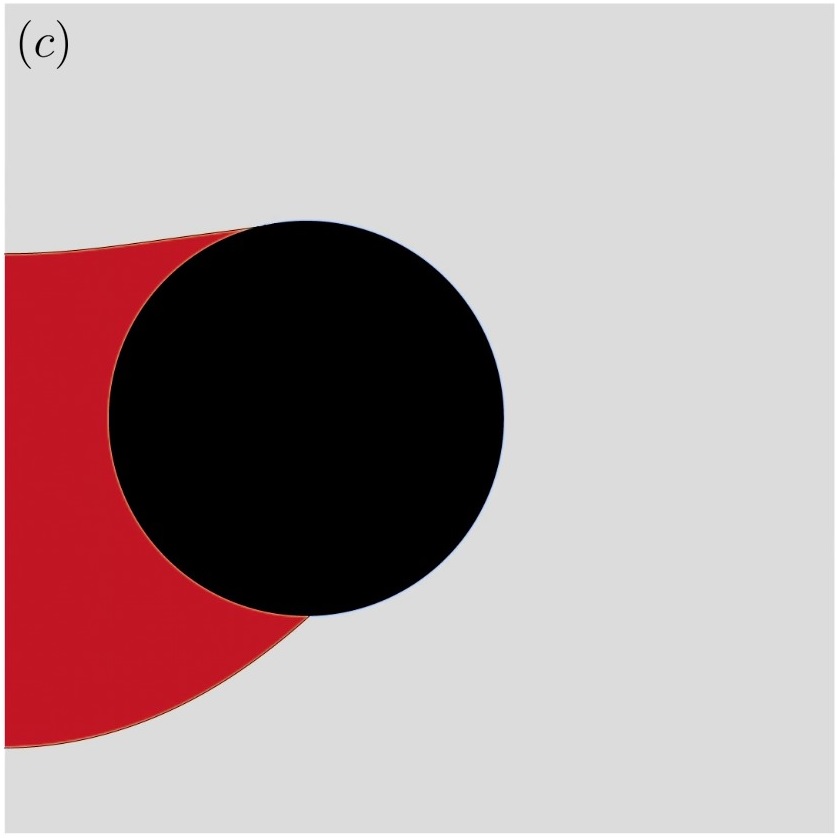}
            \includegraphics[width=0.44\textwidth]{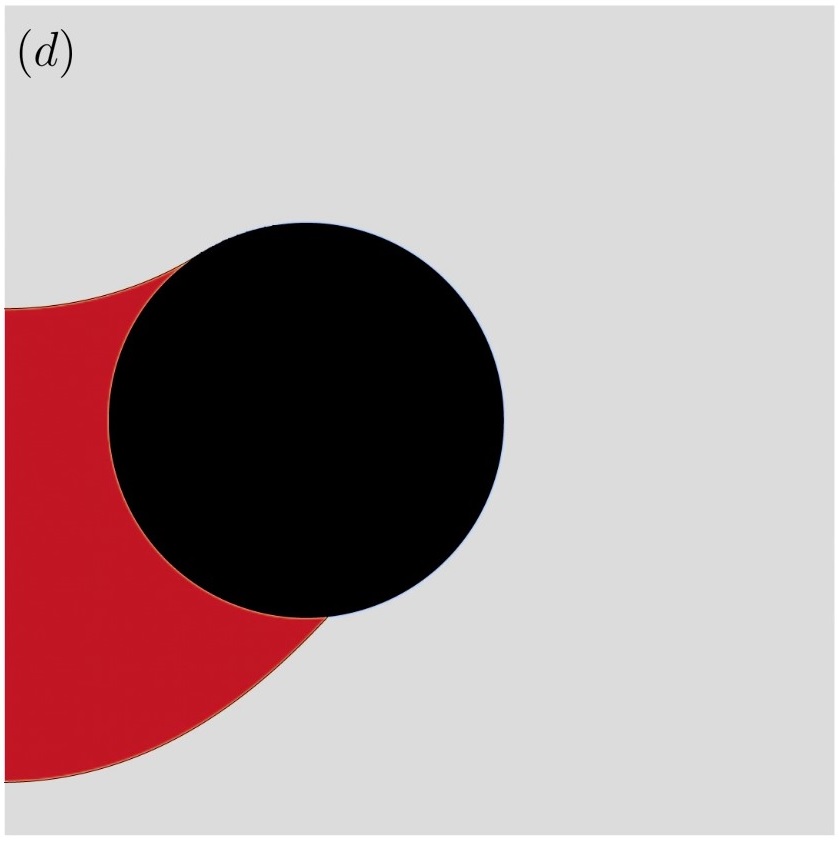}
           \caption{DNS snapshots of the liquid bridge evolution (see also associated supplementary video) for the non-electrified case of a $500\,\mathrm{cSt}$ silicone-oil bridge between copper rods, with relevant parameters given in tables~\ref{tab:Para4} and \ref{tab:Para3} (see also corresponding experiments below in \S~\ref{sec9}). The evolution is shown at dimensionless times $(a)\ t=0.0$, $(b)\ t=2.5$, $(c)\ t=5.0$ and $(d)\ t=20.0$, from the initially prescribed condition matching the equivalent reduced-order model setup towards the final stable configuration, which is compared to the results from the various other approaches in figure~\ref{DNS1}.}
           \label{DNS_TimeSeries}
     \end{figure}

\begin{figure}
    \centering
        \includegraphics[width = 0.9\textwidth]{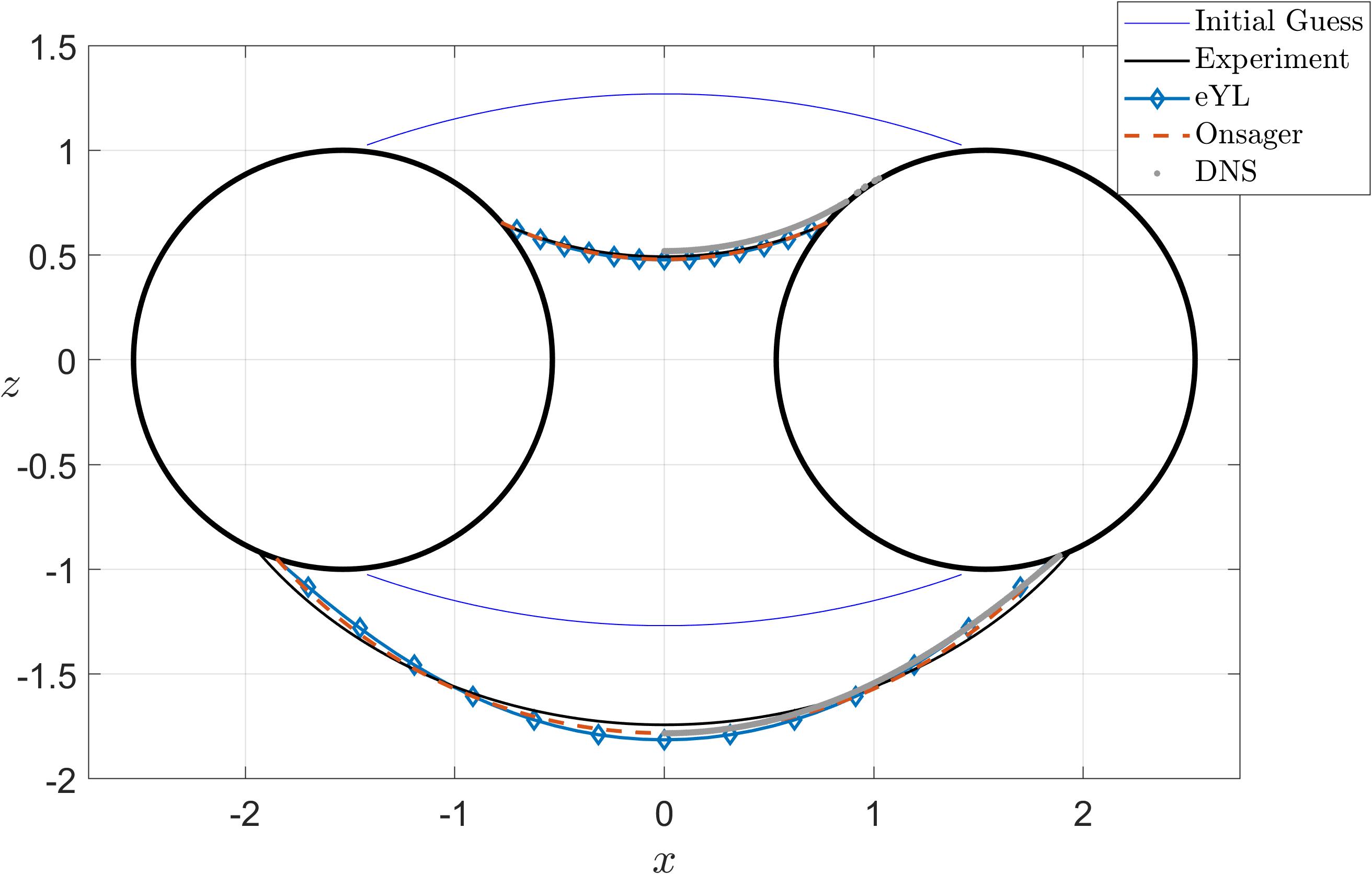}
    \caption{Comparisons of theoretical approaches with experimental data for the non-electrified case of a $500\,\mathrm{cSt}$ silicone-oil bridge between copper rods. The corresponding geometrical and dimensionless parameters are shown in Table~\ref{tab:Para3}. The physical parameters for the $500\,\mathrm{cSt}$ silicone oil are given in Table~\ref{tab:Para4}.}
    \label{DNS1}
\end{figure}

A typical numerical evolution is illustrated in figure~\ref{DNS_TimeSeries}, which shows four snapshots of the interface shape of the liquid bridge within the computational domain at the dimensionless times $t=0.0$, 2.5, 5.0 and 20.0.
These capture the key stages of the evolution within the full simulation timescale of $50.0$ time units.
Provided the viscosity is sufficiently large, the interfaces move monotonically.
The initial condition used here is identical to that used for the reduced-order Onsager model.
Almost immediately in its early time evolution, the bridge shape breaks its initial horizontal symmetry, subsequently falling under gravity towards a final state that is depicted in figure~\ref{DNS1}. This figure also shows the final state obtained from all the other techniques used in our study.
When starting with an identical initial condition as the model, we find the lower part of the bridge stops slightly short (by approximately 5$\%$ of the solid cylinder arc length) of the equivalent results obtained from the alternative techniques. 
By contrast, choosing instead an initial condition that already matches the final state at the lower contact point location (with an area-matched shape), leads to very good agreement across the entirety of the solution. The discrepancy between the observed final states arises because the contact lines become pinned at slightly different locations due to the effective roughness introduced by discretising curved solid boundaries on a structured grid. While the influence of this numerically introduced roughness can be reduced by increasing the resolution, practical limitations arise due to the computational cost of DNS in this configuration (see also the discussion in Appendix~\ref{sec:DNS}).

\begin{figure}
    \centering
        \includegraphics[width = 1.0\textwidth]{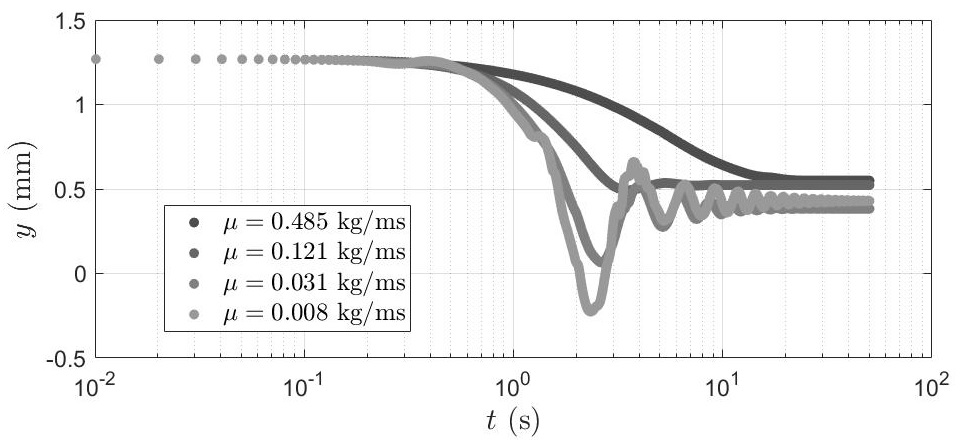}
    \caption{Position $y$ of the mid-point of the lower interface over time, for a liquid bridge relaxing towards equilibrium, for several values of the dynamic viscosity, as given in the legend. These are all for a non-electrified $500\,\mathrm{cSt}$ silicone-oil bridge between copper rods. The largest dynamic viscosity value corresponds to the case described previously in figure~\ref{DNS1}. With decreasing dynamic viscosity, we see that the evolution of $y$ undergoes a transition from monotonic to underdamped oscillatory behaviour.}
    \label{DNS2}
\end{figure}

As briefly touched upon above in our discussion of the Onsager model, the viscosity (or equivalently the dissipation matrix $\boldsymbol{D}$) sets the overall timescale for the bridge equilibration dynamics. However, when the viscosity is sufficiently low, there is a substantial change in the dynamics. 
{Namely, as the viscosity is decreased, inertial effects become significant, and the system undergoes a transition from overdamped monotonic relaxation dynamics to underdamped oscillatory behaviour that is not captured by the Onsager model.}
Figure~\ref{DNS2} shows the evolution of the position of the lowest point of the bridge $y$ for four different values of the dynamic viscosity, going down to two orders of magnitude lower compared to the experimental base case.
We see that as the dynamic viscosity is decreased, the dynamics of $y$ over time begins to exhibit an overshoot and small-amplitude oscillations as it relaxes toward equilibrium, while for even lower viscosities, a large overshoot is observed, accompanied by oscillatory decay towards equilibrium, corresponding to underdamped dynamics.
For the lowest viscosity value, secondary higher frequency oscillations can also be observed superposed upon the larger amplitude lower frequency main oscillations.
We defer the understanding of this transition from overdamped to underdamped dynamics to a future study.

\begin{figure}
    \centering
        \includegraphics[width = 0.495\textwidth]{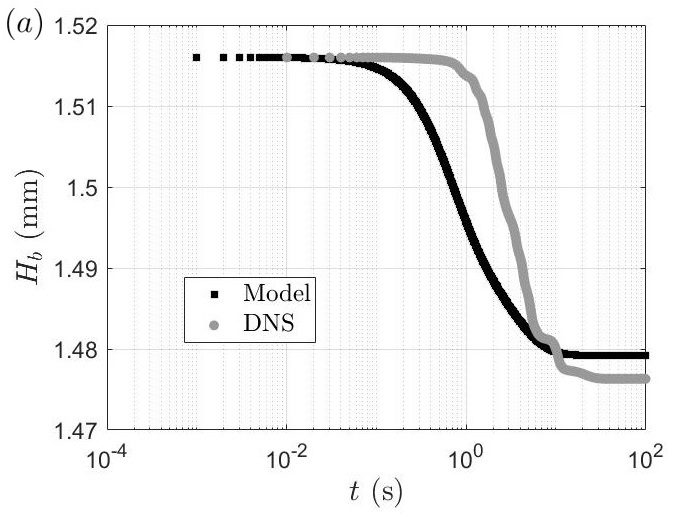}
        \includegraphics[width = 0.495\textwidth]{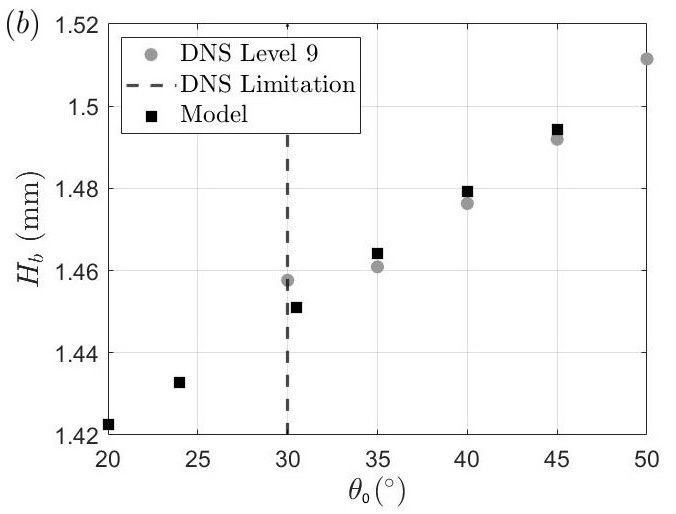}
    \caption{Panel (a) shows the variation over time of the bridge thickness $H_b$ over time, for the non-electrified case of castor oil suspended between copper rods previously presented in Figure~\ref{castor_front}, comparing the DNS result with that from the Onsager model. They both have the same initial condition and equilibrium contact angle $\theta_0 = 40^{\circ}$. Panel (b) summarises the final liquid bridge thickness results for a range of values of $\theta_0$, comparing the Onsager model results with those from the DNS (VOF) approach, which we see starts to fail for $\theta_0\lesssim 30^{\circ}$ (indicated using a dashed vertical line).}
    \label{DNS3}
\end{figure}

We must emphasise that the DNS has no fitting parameters, and as such it is a valuable calibration tool for determining the overall magnitude of the dissipation matrix $\boldsymbol{D}$ in the Onsager model. In figure~\ref{DNS3}(a), we plot the evolution of $H_b$ over time, as obtained from the DNS together with the result from the Onsager model.
For the later, in order to go from dimensionless time to the real time, we are free to adjust the overall magnitude of the dissipation matrix. Doing this, we observe good qualitative agreement.
Note that the DNS dynamics exhibits a delayed onset of the decrease in $H_b$, followed by  an indication of stick--slip-type motion. This behaviour is related to how the contact line is treated numerically. In particular, representing curved boundaries on a structured grid introduces a small effective roughness of the solid surfaces, which affects the contact line dynamics. Although this can be reduced by increasing the resolution, it may still affect the dynamics at low velocities. We also note that accurate numerical treatment of moving contact lines in VOF-type methods remains an active area of research \citep{tavares2024coupled, chen2025volume, huang20252d}, even for simpler geometrical configurations. In figure~\ref{DNS3}(b), we show how the final bridge thicknesses compare between the DNS and the Onsager model, for varying equilibrium contact angle $\theta_0$.
We observe that for the imposed equilibrium contact angle values in the range $30^{\circ} < \theta_0 < 50^{\circ}$, there is remarkably good agreement between the two approaches.
The DNS approach with embedded boundaries is known to have limits of applicability for very low or very large values of the contact angle.
For our problem, we have identified the lower limit to be approximately $30^{\circ}$ across a range of numerical experiments.

The comparisons in this section show that the Onsager model provides an accurate description of the dynamics in the overdamped regime, corresponding to sufficiently large liquid viscosity. 
Recall also that for equilibrium configurations,
the Onsager model is also in good agreement with the YL model, provided that the liquid volume is small enough (see \S~\ref{subsec:8.1}). 
Taken together, these cross-validations establish the validity of the Onsager model in the overdamped regime for moderate liquid volumes. Extending the analysis to include electric-field effects in the dynamics is left for future work.

\section{Concluding remarks}
\label{sec:10}

We have investigated the behaviour of liquid bridges suspended between two parallel horizontal cylinders, combining experiments with reduced-order modelling and DNS. We considered both non-electrified and electrified cases, applying a potential difference between the cylinders. In the experiments, (dielectric) transformer oils  were used. The results show that an applied electric field strongly modifies the bridge morphology, counteracting gravity: as the electric field strength increases, the interfaces become significantly flatter and the bridge is pulled further into the gap between the cylinders, allowing larger volumes of liquid to be supported. These trends are observed consistently in the experiments and are well reproduced by the modelling approaches.

To model the system, we simplified the full 3D problem by assuming that the liquid bridge is long and invariant in the direction parallel to the rods, allowing us to consider just the 2D cross-section case. For the static case, the problem reduces to the modified YL equations with additional contributions due to the electric field (the eYL equations), which accurately predict the observed shapes of the liquid--air interfaces and their transition towards flattening under increasing electric field strength. For the dynamics, we considered the non-electrified case and developed a reduced-order model based on the Onsager variational principle. Despite the simplifying assumption of circular interfaces, the model qualitatively reproduces the transient dynamics in the overdamped regime. Comparisons with DNS implemented within the Basilisk framework confirm this agreement, while also highlighting the breakdown of the reduced-order model at lower viscosities, where inertial effects lead to underdamped oscillations. In our experiments, we saw some evidence for this transition from underdamped oscillatory to overdamped monotonic decay dynamics, but we have not pursued this systematically and this is something we plan to study in the future.

The present framework has several limitations. The reduced-order model relies on geometric simplifications that become less accurate for larger bridge volumes, where gravity leads to deviations from circular interface shapes. In addition, both the experiments and numerical simulations are in general influenced by contact line effects and surface roughness, with the latter arising in DNS from grid discretisation effects and leading to apparent pinning behaviour; indications of contact line pinning were observed in our experiments, but not systematically investigated. Such effects are not captured by the present reduced-order model. Finally, the electrified problem has been restricted to steady states, 
with the analysis of the dynamics of the fully coupled time-dependent electrohydrodynamic problem left for future investigation.

There are several other potential avenues for future research. The reduced-order modelling could be extended to incorporate more general interface geometries, improved treatment of dissipation mechanisms, and inertial effects. Extending both the modelling and simulations to 3D configurations would enable more direct comparison with experiments. More broadly, the ability to control dielectric liquid bridges using electric fields suggests potential applications, for example, in tunable optical systems, where the bridge shape could be used to manipulate light \citep{chen2017liquid}.

\section*{Acknowledgements}

\noindent {\bf Supplementary Material.} Supplementary movies are available at [URL].

\noindent {\bf Funding.} A.J.B.-T.\ was supported by an EPSRC studentship.

\noindent {\bf Declaration of Interests.}
The authors report no conflict of interest.

\noindent {\bf Data availability.}
The DNS implementation, as well as associated data pre- and post-processing scripts are made available at \url{https://github.com/rcsc-group/LiquidBridge2D}.

\appendix 
\section{Direct numerical simulation of the non-electrified case}
\label{sec:DNS}

A direct numerical simulation implementation for the non-electrified case is created using the open-source Basilisk framework \citep{popinet2009accurate,popinet2015quadtree}, making use of state-of-the-art implementation capabilities for embedded solids \citep{tavares2024coupled} required for the geometrical features in our computational domain. The full Navier--Stokes equations, with suitable boundary and interfacial conditions, are solved in both liquid and air using a finite-volume discretisation and a VOF method to account for the multiple phases. The initial conditions are typically chosen to match those applied in the Onsager reduced-order model.

The liquid physical properties are used as reference quantities and the cylinder radius $r$ as the reference lengthscale, with a velocity scale constructed by setting the Froude number $\textrm{Fr}=1$, yielding $U = \sqrt{gr}$. All quantities are otherwise set to mimic the experimental setup, with the full computational domain measuring $3.6$ dimensionless spatial units built with an axis of symmetry in between the two parallel cylinders. The resolution level required for mesh-independent results within a rigorous tolerance level is found to be $2^9$ grid cells per dimension, amounting to approximately $140$ points per cylinder radius. We take advantage of the parallelisation features of the code, with simulations performed on local high-performance computing clusters on either $8$ or $16$ CPUs, with a typical runtime of approximately $400$ CPU hours per simulation. The quality of the results is assessed by checking liquid mass conservation and by comparing the dynamics of the liquid--air interfaces with the results obtained from the other approaches. 

The primary challenge of the DNS architecture lies in resolving the moving contact line motion on curved embedded solids. Note that our previous investigations within the same type of code architecture have been successfully applied to simpler (from a numerical standpoint) fully coated cylinder configurations \citep{wray2020reduced}.
A uniform grid is used in order to ensure result consistency and to avoid numerical artefacts being introduced by refinement processes. We stress that contact line dynamics in discretised systems is an ongoing active research area \citep{tavares2024coupled, chen2025volume, huang20252d}. 
In this context, two key methodological difficulties remain.
Firstly, the method is in practice restricted to moderate contact angles (typically $30^{\circ} < \theta_0 < 150^{\circ}$), thereby limiting access to a number of experimental and reduced-order model results in our investigation.
Secondly, the numerically-induced stick--slip motion due to the numerically-introduced effective roughness of the discretised solid surface 
at sufficiently low velocities makes the flow dynamics sensitive to the choice of initial conditions, where different initial conditions may lead to convergence to different, albeit often nearby, equilibrium configurations. In our numerical explorations, a variety of initial condition setups is therefore used to investigate convergence, accuracy and robustness of our results. 

\bibliographystyle{jfm}


\end{document}